\documentclass[journal]{IEEEtran}
\usepackage{pgfplots}
\pgfplotsset{compat=1.18}
\usepackage{colortbl}
\usepackage{amsmath,amsfonts}

\usepackage[ruled,linesnumbered]{algorithm2e}

\SetCommentSty{mycommfont}

\usepackage{array}
\usepackage{textcomp}
\usepackage{stfloats}
\usepackage{url}
\usepackage{verbatim}
\usepackage{graphicx}
\usepackage{cite}
\usepackage[colorlinks=true,
            linkcolor=blue,
            anchorcolor=blue,
            citecolor=blue,
            urlcolor=black]{hyperref}

\AtBeginDocument{%
  }

\begin{document}

\title{
    Efficient Hardware Accelerator Based on Medium Granularity Dataflow for SpTRSV
}

\author{Qian Chen, Xiaofeng Yang, and Shengli Lu,~\IEEEmembership{Member,~IEEE}
\thanks{Q. Chen, X. Yang, and S. Lu are with the Department of National ASIC System Engineering Research Center, Southeast University, Nanjing 210096, China (e-mail: chenqian0103@seu.edu.cn; lsl@seu.edu.cn).}
\thanks{This research work is supported by the Big Data Computing Center of Southeast University.}
}

\maketitle

\begin{abstract}

Sparse triangular solve (SpTRSV) is widely used in various domains. Numerous studies have been conducted using CPUs, GPUs, and specific hardware accelerators, where dataflows can be categorized into coarse and fine granularity. Coarse dataflows offer good spatial locality but suffer from low parallelism, while fine dataflows provide high parallelism but disrupt the spatial structure, leading to increased nodes and poor data reuse. This paper proposes a novel hardware accelerator for SpTRSV or SpTRSV-like DAGs. The accelerator implements a medium granularity dataflow through hardware-software codesign and achieves both excellent spatial locality and high parallelism. Additionally, a partial sum caching mechanism is introduced to reduce the blocking frequency of processing elements (PEs), and a reordering algorithm of intra-node edges computation is developed to enhance data reuse. Experimental results on 245 benchmarks with node counts reaching up to 85,392 demonstrate that this work achieves average performance improvements of 7.0$\times$ (up to 27.8$\times$) over CPUs and 5.8$\times$ (up to 98.8$\times$) over GPUs. Compared to the state-of-the-art technique (DPU-v2), this work shows a 2.5$\times$ (up to 5.9$\times$) average performance improvement and 1.7$\times$ (up to 4.1$\times$) average energy efficiency enhancement.

\end{abstract}


\section{Introduction}

\IEEEPARstart{S}{parse} triangular solve (SpTRSV) is a crucial computational kernel extensively used in various domains, including direct solvers \cite{DirectMethods, DirectMethods-2, GPU-LU, GPU-LU-2}, preconditioned iterative solvers \cite{IterativeMethods, IterativeMethods-2}, and least square problems. Different from the sparse matrix-vector multiplication (SpMV) \cite{SpMV} and sparse matrix-matrix multiplication (SpMM) \cite{SpMM}, the outputs of SpTRSV depend on the previous inputs, making it inherently sequential. It is also frequently executed in many scientific and engineering applications, such as transient simulations with fixed steps for linear circuits \cite{PGT-SOLVER, CircuitSimulation, ParallelDirect}. As a result, SpTRSV has become a performance bottleneck in many applications \cite{GPU-SpTRSV-kernel}.

{
    \begin{figure*}[t]
    \centerline{\includegraphics[width=1.0\linewidth]{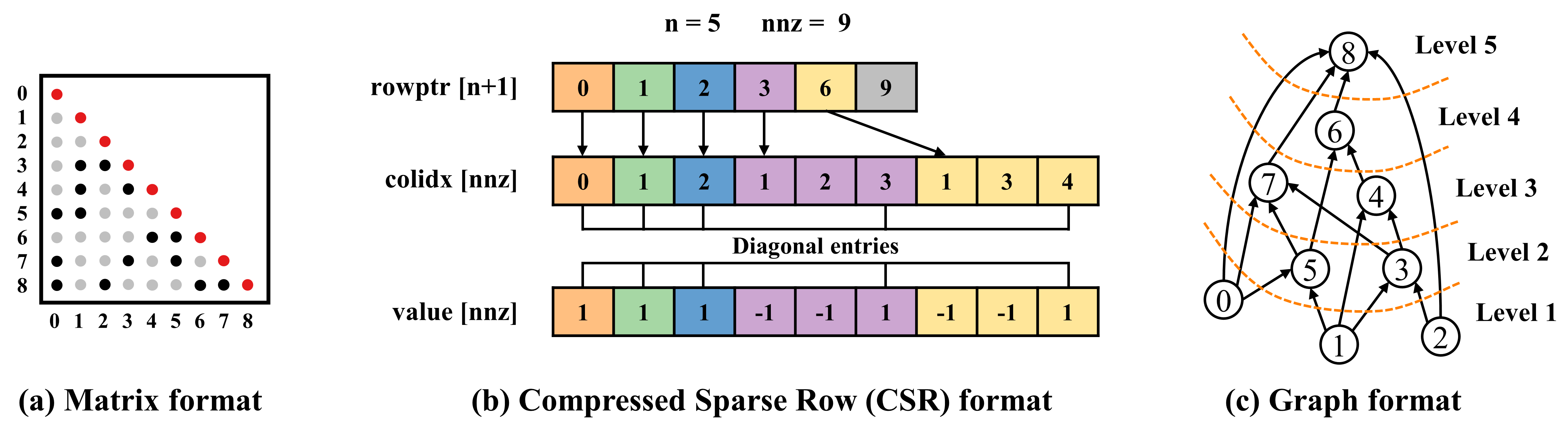}}
    \caption{
    Three formats of a sparse triangular matrix. 
    Assuming the values of diagonal entries are 1 and others are -1, the CSR format represents the first five rows of the matrix in (a). $n$ represents the matrix order and $nnz$ represents the number of non-zeros. 
    (c) also illustrates the level-scheduling method \cite{level-scheduling}.
    }
    \label{fig:matrix-format}
    \end{figure*}
}

SpTRSV is a sparse irregular graph computation represented by a directed acyclic graph (DAG), where nodes represent matrix rows and edges represent inter-row connections formed by off-diagonal non-zeros. For instance, in figure \ref{fig:matrix-format}, the node 3 represents the third row of the matrix, where column indexes of the off-diagonal non-zeros are 1 and 2, corresponding to the direct edges from node 1 and 2 to node 3 in the graph, which indicates that the computation of the third row requires the solution of the first and second rows. Since there are no circular dependencies between rows, the graph is acyclic. The level-based or node-based methods can be used to accelerate such a DAG. The level-based method, also called the level-scheduling method or level-set method, segments independent nodes into levels and utilizes multiple threads to calculate the nodes in each level simultaneously. The threads between levels need to be globally synchronized to maintain the data dependency. The node-based method, also called the synchronization-free method, allocates nodes directly to threads without building the levels and each thread commences computation once all its dependent nodes have been solved. The former is typically applied in CPUs \cite{level-scheduling, level-scheduling-2}, whereas the latter is used in GPUs \cite{sync-free-1, sync-free-2} because GPUs have more lightweight threads than CPUs. Furthermore, some studies focus on exploiting 2D spatial structure of the coefficient matrix \cite{sptrsv-spmv-1, sptrsv-spmv-2, TileSpTRSV} or reducing dependencies \cite{sptrsv-rewrite, sptrsv-rewrite-2} to optimize SpTRSV based on CPUs and GPUs.

Accelerating SpTRSV on CPUs and GPUs faces some challenges: 1) SpTRSV requires frequent synchronization, but the associated overhead on both CPUs and GPUs is substantial; 2) each node has too few edges to be worth dedicating a separate thread on either CPUs or GPUs, even when these nodes are grouped into levels; 3) the irregular sparse nature disrupts both temporal and spatial locality, leading to inefficient memory access patterns on CPUs and GPUs. Designing a specific hardware architecture can effectively tackle these challenges. For example, Nimish Shah et al. employ processing elements (PEs) and software-managed scratchpads (or register files) to replace threads and hardware-managed caches, respectively \cite{DPU-v2, DPU, DPU-book}. They refer to the design as the DAG processing unit (DPU and DPU-v2). Compared to the hardware-managed caches, the software-managed scratchpads have a lower granularity of data access and are more controllable. This allows nodes to be placed in the appropriate locations in advance based on the data dependencies, addressing the issue of irregular memory access. Furthermore, the synchronization costs in a specific architecture are significantly lower than CPUs or GPUs. To address the limited parallelism of the original DAG, they transform it into a binary DAG by cascading multiple two-input nodes to represent each multi-input node. The binary DAG is then mapped onto a tree-shaped PEs array, thereby achieving higher parallelism.

We refer to the original DAG as coarse DAG and the binary DAG as fine DAG. Nodes in the coarse and fine DAGs are called coarse nodes and fine nodes, respectively. Thus, a coarse node is a node with several inputs with multiple basic operations (e.g., addition or multiplication), while a fine node is a node with two inputs with one basic operation. A coarse (or fine) dataflow is a dataflow for a coarse (or fine) DAG, where a coarse (or fine) node is considered the minimal task scheduling unit. In this paper, the node-based and level-based methods are classified as coarse dataflows, while the the methods used by DPU and DPU-v2 are classified as fine dataflows. Although the fine dataflow improves parallelism, its performance still falls short of expectations. Compared to the coarse DAG, the fine DAG introduces numerous intermediate nodes and disrupts the spatial structure of the original matrix. In the coarse dataflow, the partial sum of a coarse node can be reused until completion. In contrast, in the fine dataflow, due to a coarse node being replaced by multiple cascaded binary nodes, the partial sum can only be written back to the register files if the cascading depth exceeds the designed PEs array. The cascading depth, which depends on the number of inputs of a coarse node, is variable and can easily reach tens or even hundreds. However, the depth of the tree-shaped PEs array is typically fixed and usually lower than ten, leading to inefficient data reuse. Additionally, the increased number of nodes exacerbates bank conflicts, further degrading performance.

In this paper, we propose a novel hardware accelerator to address the aforementioned challenges. The contributions of this paper are summarized as follows.
{
    \begin{itemize}

    \item  A novel hardware accelerator, complemented by a custom compiler, is proposed to accelerate SpTRSV or SpTRSV-like DAGs. The accelerator utilizes a medium granularity dataflow to achieve high performance. The dataflow incorporates a coarse node allocation method to maintain the spatial structure and a fine edge computation method to increase parallelism.
    
    \item  A partial sum caching mechanism is introduced to store the intermediate results of a coarse node when there are no computable edges, allowing PEs to proceed with the computation of other nodes. The mechanism reduces the blocking frequency of PEs and enhances overall performance.
    
    \item  A reordering algorithm for intra-node edges computation is proposed. The algorithm schedules the computation of similar edges within the same cycle, to enhance data reuse and reduce bank constraints and conflicts.
    
    \item  Experimental results demonstrate that, compared to the state-of-the-art technique, our approach achieves an average throughput of 6.5 GOPS (up to 14.5 GOPS), a performance speedup of 2.5$\times$, and an energy efficiency improvement of 1.7$\times$. The PEs utilization in this work can reach up to 75.3\%.
    
    \end{itemize}
}

\section{Background}

\subsection{Sparse Triangular Solve on CPUs and GPUs}

{
    \begin{algorithm}[t]
    \footnotesize
    \caption{Serial sparse triangular solve}
    \label{algo:sptrsv}
    \SetKwInOut{Input}{Input}
    \SetKwInOut{Output}{Output}
    \Input{
    $n$, $rowptr$, $colidx$, $value$: A $n\times n$ sparse triangular matrix stored in the format of compressed sparse row.\\
    $b$: The right-hand sides.
    }
    \Output{
    $x$: The solution vector.
    }
    $x \gets 0$\\
    \For{$i \gets 0$ \KwTo $n-1$}{
        $ie \gets rowptr[i+1]-1$\\
        $sum \gets 0$\\
        \For{$j \gets rowptr[i]$ \KwTo $ie-1$}{
            $sum \gets sum + value[j] \times x[colidx[j]]$
        }
        $x[i] \gets (b[i] - sum) / value[ie]$
    }
    \Return{$x$}
    \end{algorithm}
}

The sparse triangular solve can be represented by solving $Lx=b$, where $L$ is a sparse triangular matrix and $b$ is the right-hand sides (RHS). Sparse matrices are commonly stored in formats like compressed sparse row (CSR) and compressed sparse column (CSC) \cite{CSR-CSC}. In CSR format, a sparse matrix is represented by three vectors, as shown in figure \ref{fig:matrix-format} (b). one stores the position indices of the first non-zero in each row ($rowptr$), another stores the column indices of non-zeros in row-major order ($colidx$), and the third vector stores the corresponding values ($value$). The number of non-zeros ($nnz$) is placed at the end of the vector $rowptr$ and the diagonal entries are placed at the end of each row (line 3 in algo. \ref{algo:sptrsv}). The CSC format is similar but stores entries in column-major order. A basic approach to solving the triangular system is elimination. The triangular matrix shows that the unknowns in the solution vector are influenced only by preceding unknowns. Thus, the elimination can be performed sequentially based on the original order of the solution vector. Assuming the first $j-1$ rows are solved, solving for the $j$-th unknown ($x_j$) involves using the column indices of the non-zeros in this row to index the corresponding entries in the solution vector and perform the multiply-accumulate operations (lines 5-7 in algo. \ref{algo:sptrsv}). The RHS $b_i$ and the diagonal entry $L_{ii}$ are then used to update and obtain the final result, as shown in equation \ref{sptrsv} or line 8 in algo. \ref{algo:sptrsv}.

{
    \begin{small}
    \begin{equation}
    L x=b \hspace{1em} \Rightarrow \hspace{1em} x_i=\frac{1}{L_{i i}}\left(b_i-\sum_{j=1}^{i-1} L_{i j} x_j\right)
    \label{sptrsv}
    \end{equation}
    \end{small}%
}

For sparse matrices, calculating each unknown does not utilize all preceding results, allowing for mining the parallelism. Sparse triangular matrices can be converted into DAGs, where each edge represents a multiply-accumulate operation and each node represents an unknown and a self-update operation (subtraction and division). The level-scheduling method \cite{level-scheduling} divides all nodes into multiple levels, where nodes within the same level are independent and have the same depth from the source node. The nodes within the same level can be computed simultaneously by multiple threads, but synchronization between levels is necessary to ensure data dependencies, as shown in figure \ref{fig:matrix-format} (c). The synchronization-free method \cite{sync-free-2} assigns a node to a warp on GPUs and begins parallel computation for the input edges after all dependent nodes have been solved, without waiting for other nodes at the same level. For example, in figure \ref{fig:matrix-format} (c), node 4 can enable computation once nodes 1 and 3 have been solved, without waiting for node 5. This method achieves higher parallelism but results in more communication overhead. In addition, the block and tiled algorithms \cite{sptrsv-spmv-1, sptrsv-spmv-2, TileSpTRSV} are proposed to exploit intra-node parallelism, and the rewrite method \cite{sptrsv-rewrite} is introduced to reduce DAG dependencies, to achieve better performance.

However, using CPUs and GPUs to accelerate SpTRSV remains challenging \cite{FPGA-SpTRSV, FPGA-SpTRSV-2}. On one hand, the sparse irregular data distribution disrupts temporal and spatial locality, rendering high-access-granularity caches inefficient \cite{challenge-cpu-gpu, challenge-cpu-gpu1, challenge-cpu-gpu2, challenge-cpu-gpu3}. For example, in modern CPUs and GPUs, the minimum data width transferred from the memory to cache is typically 32 or 64 words. Due to irregular distribution, only one of these words may be used, causing high cache misses and frequent data transfers. On the other hand, SpTRSV requires frequent synchronization, and the computation per node is minimal, leading to threads on CPUs and GPUs consuming most time in communication rather than computation. Therefore, the customizable hardware platform may be an attractive solution to accelerate SpTRSV.

\subsection{Very-Long-Instruction-Word Architecture}

The very-long-instruction-word (VLIW) architecture is a design approach for parallel processors. It enhances parallelism and processing efficiency by merging multiple operations into a single instruction word and executing them simultaneously \cite{VLIW-1, VLIW-2, VLIW-3, VLIW-4, VLIW-5}. Unlike traditional architectures, which rely on hardware for instruction scheduling and parallel execution, the VLIW architecture utilizes the compiler to explore instruction-level parallelism. This approach reduces the complexity of control logic and simplifies hardware design. By integrating software-managed register files, the VLIW architecture can implement communication between compute units (CUs) without synchronization overhead. Therefore, the VLIW architecture is well-suited for accelerating DAG tasks that require complex scheduling logic and frequent synchronization, such as SpTRSV. The main disadvantage of the VLIW architecture is that the compiler becomes more complex.

\subsection{DAG Processing Unit and This Work}

{
    \begin{figure}[t]
    \centerline{\includegraphics[width=1.0\linewidth]{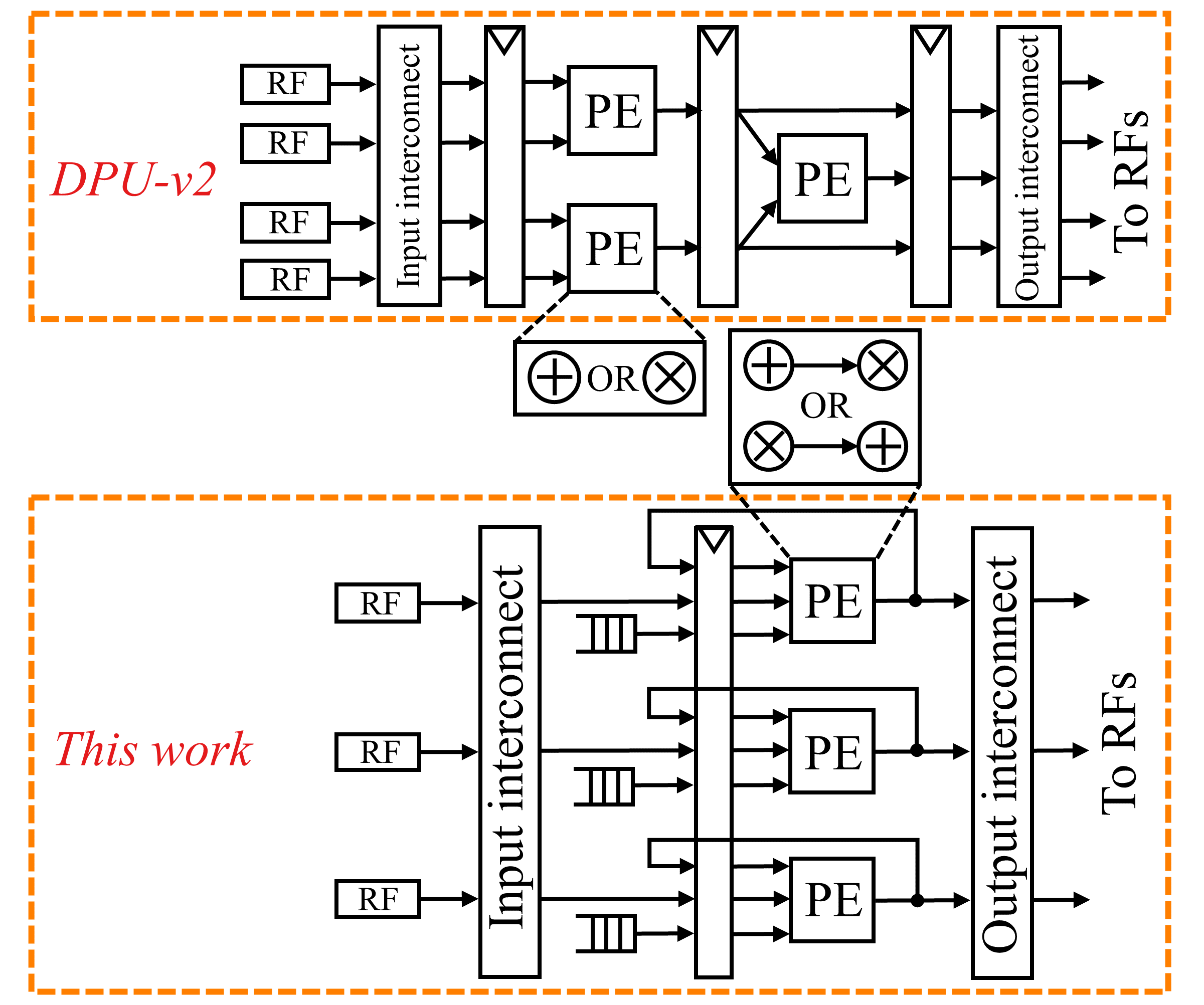}}
    \caption{
    Architecture comparison of DPU-v2 \cite{DPU-v2} and this work.
    }
    \label{fig:architecture-comparison}
    \end{figure}
}

In recent years, a few studies have attempted to accelerate SpTRSV based on customized architectures. The latest and representative work is the DAG processing unit v2 (DPU-v2) \cite{DPU-v2}. Based on the VLIW architecture, DPU-v2 utilizes tree-shaped PEs arrays and pipelining techniques to enhance data reuse. Each PE has two inputs and one output, performing a basic operation (e.g., addition or multiplication), as illustrated in figure \ref{fig:architecture-comparison}. To map such arrays, the DPU-v2 compiler converts the coarse DAG into the fine DAG by inserting multiple cascaded binary nodes. This approach theoretically maximizes parallelism since a coarse node is divided into multiple fine nodes that can be computed simultaneously. While DPU-v2 targets general DAG tasks, it is only effective for DAGs where each edge represents a single basic operation and each node contains few edges. DAGs from machine learning, like sum-product networks \cite{sum-product, sum-product-2}, often meet these conditions.

{
    \begin{figure}[t]
    \centerline{\includegraphics[width=0.9\linewidth]{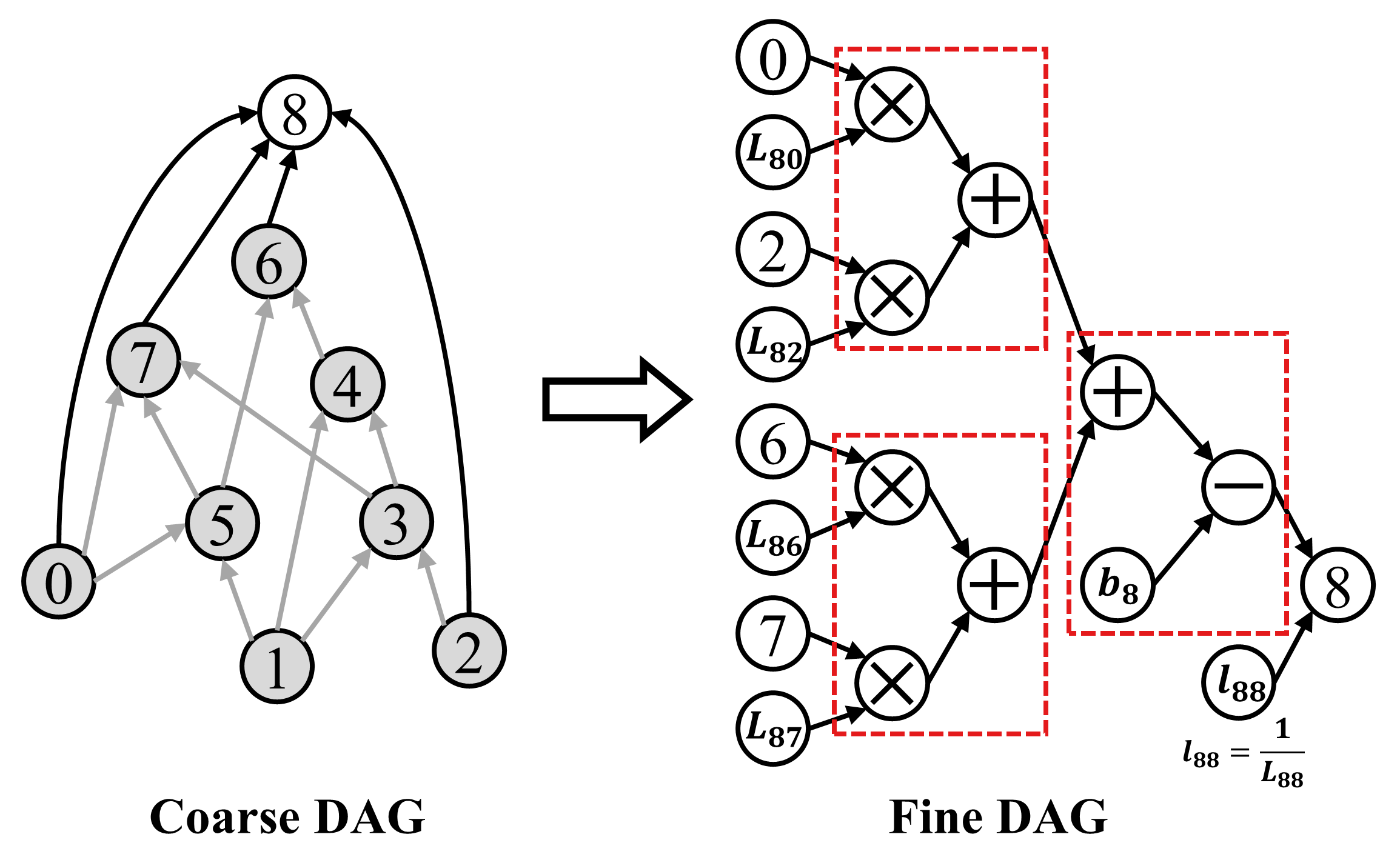}}
    \caption{
    An example of converting a coarse node 8 into multiple fine nodes and mapping them to the tree-shaped PEs, assuming other nodes have been solved. $L_{ij}$ represents the non-zeros in row $i$ and column $j$ of the coefficient matrix.
    }
    \label{fig:coarse-binary}
    \end{figure}
}

However, the DPU-v2 architecture becomes inefficient for SpTRSV-like DAGs. These DAGs have the following three characteristics: 1) each edge represents two or more cascaded basic operations, such as multiply-accumulate; 2) a node possibly contain multiple edges (more than 10); 3) most edges are concentrated on coarse-dataflow-unfriendly (CDU) nodes. In this paper, CDU nodes are defined as nodes where the number of nodes within the same level is less than a specified threshold, since the parallelism of coarse dataflows is significantly limited when processing these nodes. The first two characteristics increase the number of fine nodes and reduce data reuse of the tree-shaped PEs array. For instance, consider using a tree-shaped PEs array with a depth of 2 to calculate a coarse node with 4 input edges, where the source nodes of these edges have been solved, as shown in figure \ref{fig:coarse-binary}. The PEs array needs to be mapped 4 times to solve this node, and the intermediate results must be written back to register files in the first 3 mappings. Despite the appearance that 3 PEs are calculating a coarse node simultaneously, the pipelining structure and register files access degrade performance, potentially making it less efficient than using a single PE. Increasing the depth of tree-shaped PEs array can mitigate this issue but exponentially increases hardware complexity. The third characteristic limits the parallelism of traditional coarse dataflows, such as level-scheduling and synchronization-free methods.

This work proposes an architecture with feedback-structured PEs to achieve adaptive and efficient data reuse. The compiler determines the reuse count by calculating nodes indegree. The functionality of PE aligns with basic operations of the DAG tasks. This paper evaluates the performance of the architecture for SpTRSV. Given that the basic operations of SpTRSV are a cascade of multiplication and addition (see line 6 and 8 in algo. \ref{algo:sptrsv}, where subtraction and division can be transformed into addition and multiplication by the compiler), and that the coefficient matrix $L$ and RHS $b$ are read-only, we design the architecture as shown in figure \ref{fig:architecture-comparison}, in which PEs are assigned coarse nodes, $L$ and $b$ are fed to PEs through FIFOs, and the partial sums are reused through feedback structures. The architecture employs a novel dataflow, called the medium granularity dataflow, to overcome the parallelism limitations of coarse dataflows when handling CDU nodes. The medium granularity dataflow is defined as a dataflow for a coarse DAG, where a coarse node is considered the minimal load allocating unit and an edge is considered the minimal task scheduling unit.

\section{System Architecture}

{
    \begin{figure*}[t]
    \centerline{\includegraphics[width=1.0\linewidth]{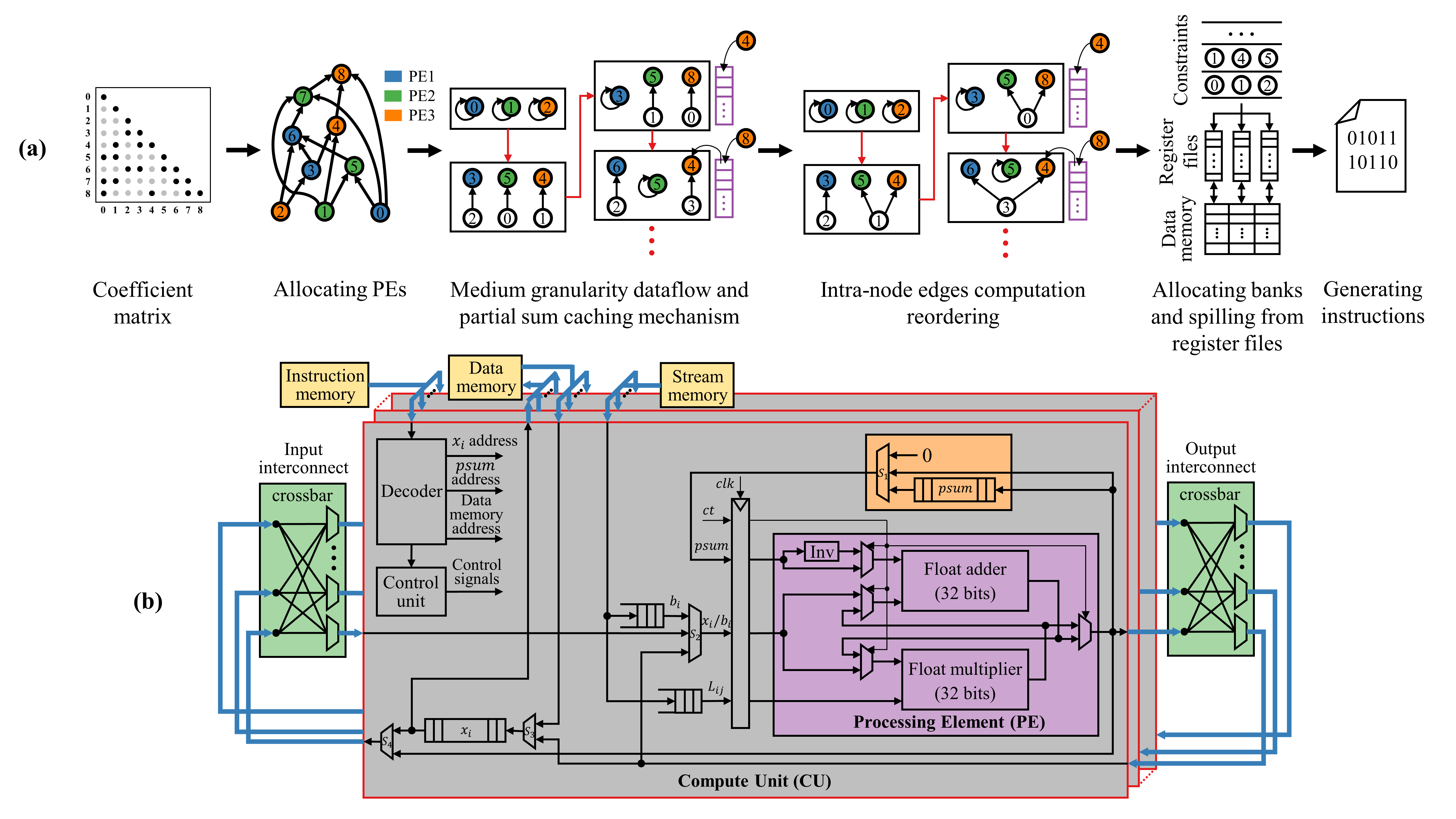}}
    \caption{
    (a) Overview of the custom compiler. The spin arrow indicates updating the node with the RHS $b$. 
    (b) The architecture of the proposed accelerator. It consists of $\text{2}^N$ compute units (CUs) connected by input and output interconnects, where $N$ is a hyperparameter. Stream memory stores the sparse matrix non-zeros ($L$) and RHS ($b$), sequentially supplying the data to CUs. Data memory stores the solution vector $x$.
    }
    \label{system-architecture}
    \end{figure*}
}

In real-world applications, a sparse triangular system is usually solved multiple times with the same coefficient matrix. We can preprocess the data structure before SpTRSV is executed because the preprocess time can be amortized. Figure \ref{system-architecture} illustrates the system architecture, comprising a software compiler and a hardware accelerator. The compiler preprocesses a sparse triangular matrix based on the proposed dataflow, maps nodes to register files, and generates instructions. The accelerator executes these instructions to obtain the solution vector and stores it in the data memory.

\subsection{Compiler Overview}

The compiler workflow is as follows: firstly, traverse the adjacency graph of the coefficient matrices and allocate nodes to PEs according to the topological order of the graph. Next, apply the medium granularity dataflow (section IV. A) and the partial sum caching mechanism (section IV. B) to establish the initial computation pattern for PEs. Then, without changing the computation order of coarse nodes and the access mode of the $psum$ register files determined in the previous step, rearrange the computation order of the edges within the nodes in each cycle (section IV. C) to enhance data reuse and reduce bank conflicts. Subsequently, determine the memory access requirements of the PEs, set constraints for nodes, and resolve bank conflicts by a greedy graph coloring algorithm. Finally, address the potential spilling issues of the register files and generate the instructions.

\subsection{Accelerator Architecture}

The proposed accelerator achieves flexible dataflow by executing the instructions, as shown in figure \ref{system-architecture} (b). It consists of multiple CUs, interconnects, and on-chip memories. The CUs are connected by input and output interconnects, which are implemented by the crossbars, enabling communication between them. The crossbar-based interconnects decouple the PEs and register files, minimizing bank conflicts. The on-chip memories include instruction memory, data memory, and stream memory. The instruction memory stores instructions generated by the compiler. The compiler is only re-executed and the instruction memory is updated when the structure of the coefficient matrix changes. The stream memory stores the values of the coefficient matrix and the RHS, without storing their positional information (i.e., in which rows or columns). This is because the data in the stream memory is reordered by the compiler through the analysis of the established medium granularity dataflow, so that the positional information is hidden in the instructions. Consequently, during the running of the accelerator, data in the instruction memory and stream memory is accessed sequentially. The accelerator writes the computed outputs to the data memory, which can be read when the register file spills.

{
    \begin{small}
    \begin{equation}
    out =\left\{\begin{array}{l}
    \vspace{0.5em}
    \left(b_i-psum\right) \times L_{ij}(ct=0) \\
    psum+L_{ij} \times x_i (ct=1)
    \end{array}\right.
    \label{pe}
    \end{equation}
    \end{small}%
}

In our architecture, the CU is a basic unit that undertakes the task of solving a coarse node, including instruction decoding, data calculation, and data caching. The PE, as an internal component of the CU, is a module for data calculation. Each CU comprises a decoder, a control unit, a D type flip-flop (DFF), an $x_i$ register file, a $psum$ register file, a PE, and several multiplexers. The PE consists of a cascaded 32-bit floating-point adder and multiplier, implementing the two fundamental operations of SpTRSV via a 1-bit control signal ($ct$ in figure \ref{system-architecture} (b)). The functionality of SpTRSV and PE are described in equations \ref{sptrsv} and \ref{pe}, respectively. The division operation is performed by computing the reciprocal in the compiler and using multiplication in the hardware. The PE output can follow two paths: one as a partial sum, which is either written to the $psum$ register file or fed back to the $psum$ input of the PE, depending on whether the current node is blocked in the next cycle. When the node in the $psum$ register file is unblocked, the partial sum is read out and the address is released. This will be discussed in detail in section IV. B. When the PE or CU starts computing a new node, the $psum$ input is set to zero. The other path is the solution for the node, which is written through the output interconnect to the $x_i$ register file of any CU, or directly reused by other PEs through input and output interconnects.

{
    \begin{figure}[t]
    \centerline{\includegraphics[width=1.0\linewidth]{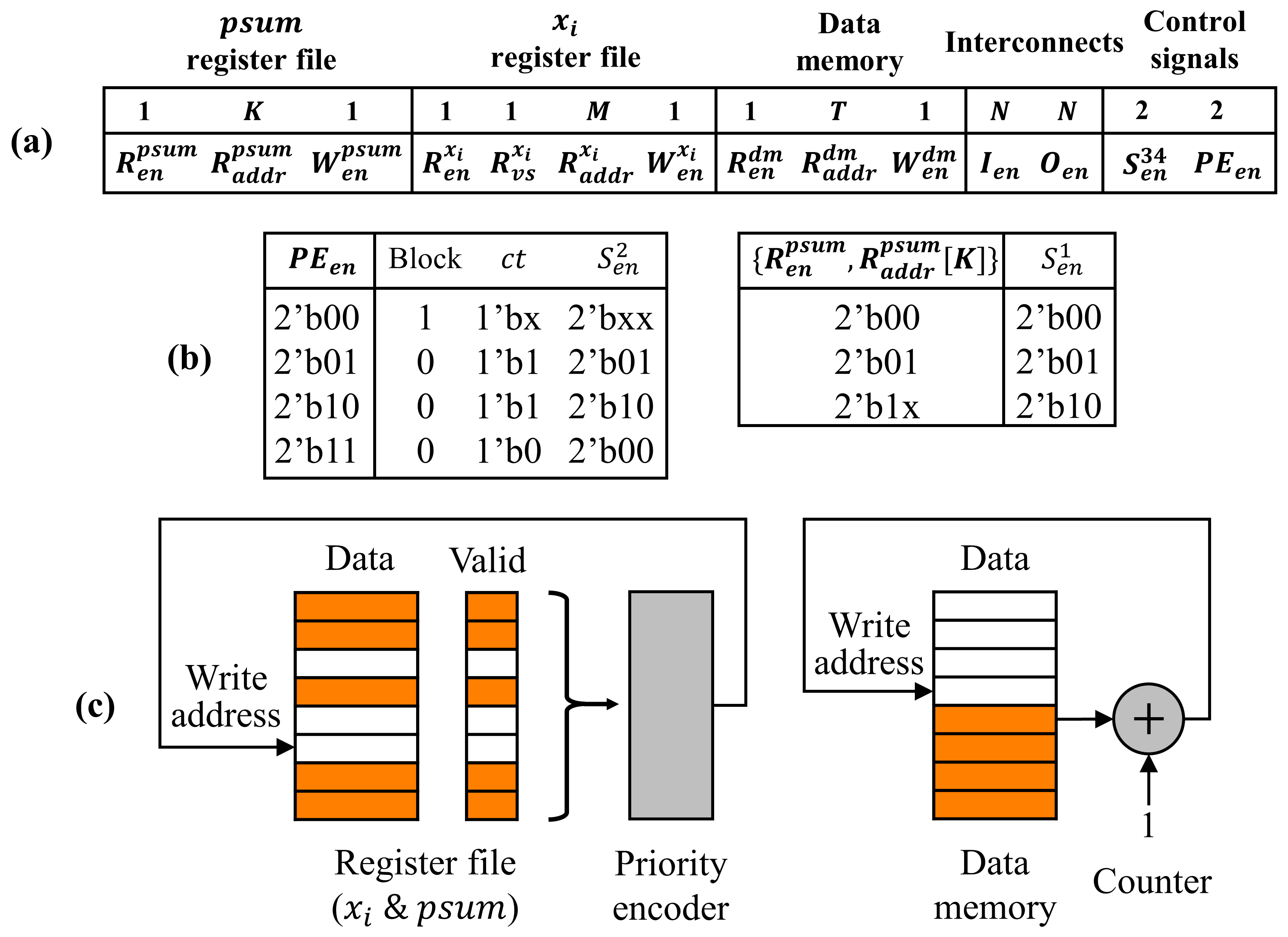}}
    \caption{
    (a) The structure and length of the instruction for each CU. Assuming that there are $\text{2}^N$ CUs, with each CU having $\text{2}^M$ and $\text{2}^K$ words in the $x_i$ and $psum$ register files respectively. Each CU has an addressing depth of $\text{2}^T$ for the data memory.
    (b) The encoding details of control signals. The definitions of relevant symbols are shown in table \ref{tab:symbol}.
    (c) Automatic write address generation for the register file and data memory.}
    \label{instruction-structure}
    \end{figure}
}

{
    \begin{table*}[htbp]
    \caption{Symbol Definitions Related to the Instruction Shown in Figure \ref{instruction-structure}}
    \begin{center}
    \renewcommand{\arraystretch}{1.5}
    \begin{tabular}{r|lcl}

    \hline
     Id & Symbol & Data width & Definition \\
    \hline
     1 & $R/W^{psum}_{en}$ & 1 & The read or write enable signals for the $psum$ register file. \\
     2 & $R^{psum}_{addr}$ & $K$ & The read address for the $psum$ register file. \\
     3 & $R/W^{x_i}_{en}$ & 1 & The read or write enable signals for the $x_i$ register file. \\
     4 & $R^{x_i}_{addr}$ & $M$ & The read address for the $x_i$ register file. \\
     5 & $R^{x_i}_{vs}$ & 1 & The control signal that determines whether releasing the read address of the $x_i$ register file in the current cycle. \\
     6 & $R/W^{dm}_{en}$ & 1 & The read or write enable signals for the data memory. \\
     7 & $R^{dm}_{addr}$ & $T$ & The read address for the data memory. \\
     8 & $I/O_{en}$ & $N$ & The control signals for the input or output interconnects.\\
     9 & $S^{34}_{en}$ & 2 & The control signals for the multiplexers $S_3$ and $S_4$ shown in figure \ref{system-architecture} (b).\\
    10 & $PE_{en}$ & 2 & The control signals for the PE and multiplexer $S_2$ shown in figure \ref{system-architecture} (b).\\
    11 & $S^{1}_{en}$ / $S^{2}_{en}$ & 2 & The control signals for the multiplexers $S_1$ or $S_2$ shown in figure \ref{system-architecture} (b). \\
    12 & $ct$ & 1 & The control signal that determines the PE calculation mode. \\
    13 & $Block$ & 1 & The signal that determines if the CU or PE is blocked. \\
    14 & $N/M/K/T$ & - & The parameters that determine the number of CUs and the capacity of the register files and data memory. \\
    \hline
    
    \end{tabular}
    \label{tab:symbol}
    \end{center}
    \end{table*}
}

The structure of the instruction for each CU is shown in figure \ref{instruction-structure} (a). The definitions of relevant symbols are shown in table \ref{tab:symbol}. The decoder translates instructions to obtain read addresses ($R^{psum}_{addr}$, $R^{x_i}_{addr}$, $R^{dm}_{addr}$) and read-write enable signals ($R/W^{psum}_{en}$, $R/W^{x_i}_{en}$, $R/W^{dm}_{en}$) for the register files and data memory, as well as selection signals ($I_{en}$, $O_{en}$) for input and output interconnects, and sends other data to the control unit. The symbol $S^{34}_{en}$ controls multiplexers $S_3$ and $S_4$ in figure \ref{system-architecture} (b). The symbol $PE_{en}$ encodes the CU blocking information and control signals of PE and multiplexer $S_2$, as shown in figure \ref{instruction-structure} (b). When the CU updates a coarse node using the RHS, the control signal $ct$ for the PE is set to 0 and $S_2$ selects the RHS data. When the CU updates a coarse node using another node, $ct$ is set to 1 and $S_2$ selects data from the interconnects. The control signal of multiplexer $S_1$ is encoded into the read enable signal ($R^{psum}_{en}$) and read address data ($R^{psum}_{addr}$) of the $psum$ register file. This is because when $R^{psum}_{en}$ is active, $S_1$ will select the $psum$ register file. When $R^{psum}_{en}$ is inactive, since the symbol $R^{psum}_{addr}$ is free, the highest bit of the address is reused to determine the control signal of multiplexer $S_1$.

Write addresses for the register files and data memory are automatically generated by the control unit to reduce instruction length. The principle for generating write addresses is to always write data to the empty location with the lowest address. The method for generating write addresses is illustrated in figure \ref{instruction-structure} (c). For the register files, a 1-bit valid flag is set for each address to determine if it is free, and a priority encoder is used to generate the write address. Whether an address in the $x_i$ register file is released after being read is determined by a 1-bit signal ($R^{x_i}_{vs}$) in the instruction. Data in the $psum$ register file is released once read out, as it is intermediate data and will not be reused by other PEs. For the data memory, write addresses are generated by a counter initialized to zero, incrementing with the write enable signal valid. This is because data written to the data memory is the final result required by the user and will not be overwritten.

All CUs in the accelerator share a single clock, and PEs achieve synchronized computation through DFFs. Thus, with a software-managed memory system, the compiler can fully predict the behavior of the hardware and the details of data stored in the register file and memory, to address the problem of bank conflicts. Data will be spilled from the $x_i$ register file if the number of nodes exceeds capacity. Given the execution schedule, a live-range analysis is performed to determine when the spilling is required. If the number of free addresses falls below a threshold, the data in the register file will be written to the data memory, or the address will be directly released if the data memory already holds the same data. The spilled data will be loaded back before it is supposed to be consumed. The compiler decides whether to insert nop cycles to perform the store and load operations by considering the port occupancy of the $x_i$ register file.

\section{Custom Compiler Methodology}

In this section, we introduce the medium granularity dataflow, the partial sum caching mechanism, and the intra-node edges computation reordering algorithm. The medium granularity dataflow combines the advantages of coarse and fine dataflows to enhance performance for SpTRSV or SpTRSV-like DAG tasks. The partial sum caching mechanism provides temporary storage to reduce PEs blocking frequency and shorten the critical path length. The intra-node edges computation reordering algorithm exploits the spatial locality to improve data reuse and alleviate constraints on banked register files.

\subsection{Medium Granularity Dataflow}

In coarse dataflow, a coarse node represents the minimal task scheduling unit. A PE or thread only starts computing a node after all its predecessors have been solved. This method is typically used in CPUs and GPUs, where the task granularity must be large enough to offset thread management overheads (e.g., startup, termination, synchronization). However, real application DAGs often have many CDU nodes, resulting in nearly sequential computation (see the coarse dataflow example in figure \ref{data-flow}). In fine dataflow, a coarse node is divided into multiple cascaded two-input fine nodes, each representing a multiplication or addition and mapped to a PE. This allows multiple PEs to compute a coarse node simultaneously. However, SpTRSV-like DAGs have two key characteristics: 1) the fundamental operations are cascaded multiplications and additions, needing two fine nodes per coarse node; 2) numerous coarse nodes with multiple inputs create many intermediate fine nodes. The two characteristics significantly increase node count and critical path length. Although a tree-shaped PE array can improve data reuse, the reuse depth is limited and many intermediate nodes still need to be written back to the register files for later use. Furthermore, the increased intermediate nodes also exacerbate bank conflicts.

{
    \begin{figure}[t]
    \centerline{\includegraphics[width=1.0\linewidth]{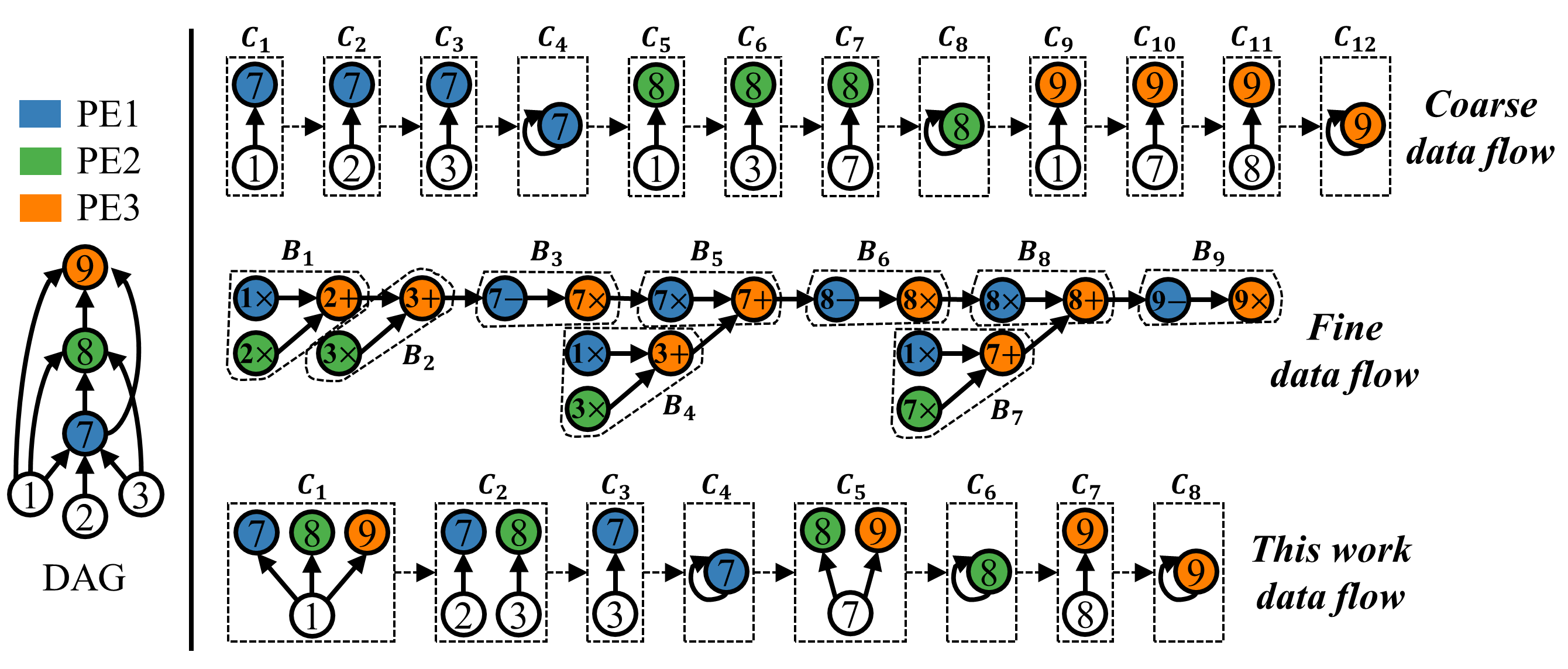}}
    \caption{Comparison of the coarse, fine, and this-work dataflows for an SpTRSV-like DAG, with nodes 1, 2, and 3 already computed. In the coarse and this-work dataflows, nodes 7, 8, and 9 are assigned to PE 1, PE 2, and PE 3, respectively. In the fine dataflow, the three PEs are connected in a tree topology \cite{DPU-v2}, and the coarse DAG is converted to the fine DAG and mapped to the PEs.}
    \label{data-flow}
    \end{figure}
}

The proposed dataflow combines the advantages of coarse and fine dataflows. Here, a coarse node is mapped to a PE, which processes any available edges immediately once its source node has been solved, without waiting for all input edges. Thus, this method can be seen as coarse allocation with fine computation. The output of a PE is fed back to the $psum$ input to reuse partial sums (orange box in figure \ref{system-architecture} (b)). Upon node completion, the output is written to the register files and routed through the interconnects for other PEs. This feedback structure adjusts the depth of data reuse based on the number of edges for each node, ensuring the partial sum is always reused or stored in the $psum$ register file, thus reducing access requirements on the register files and interconnects. The medium granularity dataflow enables parallel computation between nodes, necessitating more synchronization than the coarse or fine dataflow. This is one of the reasons that we utilize synchronized PEs based on the VLIW architecture, because the asynchronous computational architecture incurs additional synchronization costs.

Figure \ref{data-flow} illustrates an example to demonstrate how the three dataflows handle the CDU nodes. For the coarse dataflow, a PE must ensure that all dependent nodes have been solved before computing a node. This results in serial computation for the CDU nodes. In the given example, assuming no bank conflicts and the $L$ and $b$ are previously stored in register files, the coarse dataflow consumes 12 cycles. For the fine dataflow, the original DAG is converted into a binary DAG and mapped onto tree-shaped PEs. In this example, the binary DAG is divided into 9 blocks. Considering the impact of register file access and pipeline structure, these 9 blocks consume 19 cycles. Since PEs in the fine dataflow perform a single logical operation while the coarse and this-work dataflows perform 2 logical operations. For a fair comparison, we assume the fine dataflow operates at twice the clock frequency of the others. Thus, the fine dataflow consumes 9.5 cycles. In the proposed dataflow, we retain the node allocation strategy of the coarse dataflow, assigning one coarse node to a PE, but do not treat it as a minimal task scheduling unit. The PEs will compute a node if one of its dependent nodes has been solved. For the given example, this work dataflow consumes 8 cycles. In summary, the proposed dataflow demonstrates superior performance compared to both the coarse and fine dataflows.

\subsection{Partial Sum Caching Mechanism}

{
    \begin{figure}[t]
    \centerline{\includegraphics[width=1.0\linewidth]{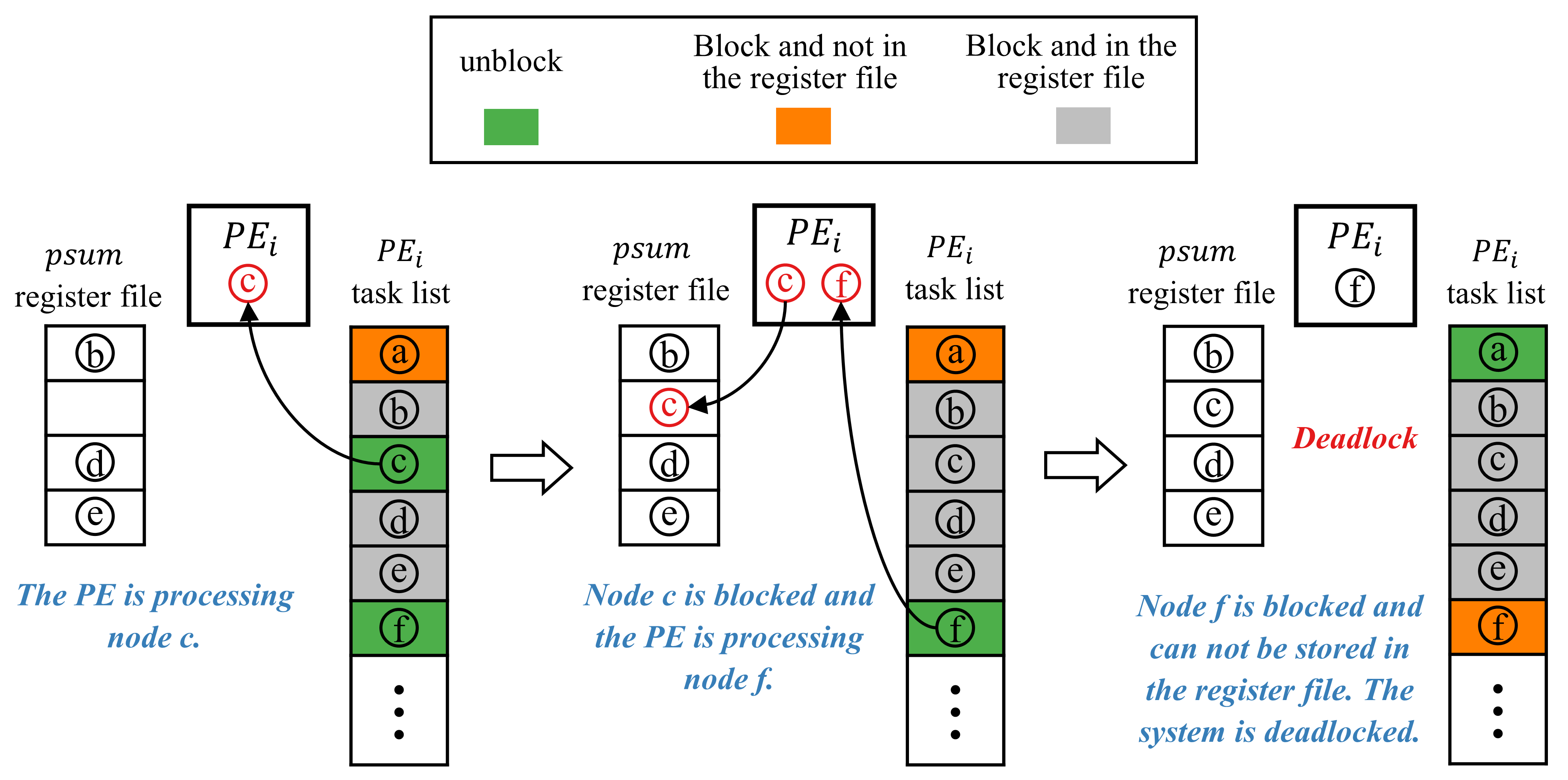}}
    \caption{An example to illustrate the deadlock.}
    \label{deadlock}
    \end{figure}
}

Due to coarse node allocation, the medium granularity dataflow may experience blocking when a PE or CU has no edges to compute for the current node. 
Blocking occurs in two scenarios: one where all nodes in the task list of the PE are blocked, and another where other nodes still have computable edges. The first scenario, related to the DAG structure and node allocation order, is rare and can be addressed by optimizing the dataflow, such as using a reconfigurable dataflow. For the more common second scenario, we introduce a partial sum caching mechanism to mitigate the blocking. In our architecture, when blocking occurs, each CU uses a $psum$ register file to cache partial sums, then traverses the task list to find the first unblocked node and calculates it. The $psum$ register file is local and does not use interconnect resources. Nodes in the $psum$ register file must be prioritized to avoid potential deadlocks, as they are always parent nodes to others. Thus, in each cycle, if a node in the $psum$ register file is unblocked, the PE must pause the current task to compute this node, regardless of whether the current node is blocked.

Determining when to perform read/write operations on the $psum$ register file is crucial. Firstly, without considering the capacity of the $psum$ register file, the task of a PE for the current cycle is determined based on the above principles. If the PE is blocked (this blocking is from the first scenario), or the current node and the previous node are the same, or the previous node has been solved and the current node is new, the PE will reuse the partial sum or set it to zero without accessing the $psum$ register file. If the previous node has been solved and the current node is old, the PE will read the partial sum from the $psum$ register file and release the corresponding address. If the previous node has not been solved and the current node is new, the PE will put the previous partial sum into the $psum$ register file and set the current partial sum to zero. In this case, the $psum$ register file must have at least two free addresses, or one free address if the current node is the first new node in the task list. Otherwise, the PE will be blocked to avoid the potential deadlock, as shown in figure \ref{deadlock}. If the previous node has not been solved and the current node is old, the PE will put the previous partial sum into the $psum$ register file and read the current partial sum from it. This scenario does not need to consider the capacity of the $psum$ register file, as it supports read-before-write operations. Figure \ref{system-architecture} (a) shows a simple example to illustrate the partial sum caching mechanism, where PE 3 is blocked while computing node 5. In the third cycle, the PE stores node 5 into the $psum$ register file and computes the next node (node 9) in its task list. Then, in the next cycle, node 5 is unblocked and read out for computation. Meanwhile, node 9 is stored in the $psum$ register file.

\subsection{Intra-node Edges Computation Reordering Algorithm}

{
    \begin{figure}[t]
    \centerline{\includegraphics[width=1.0\linewidth]{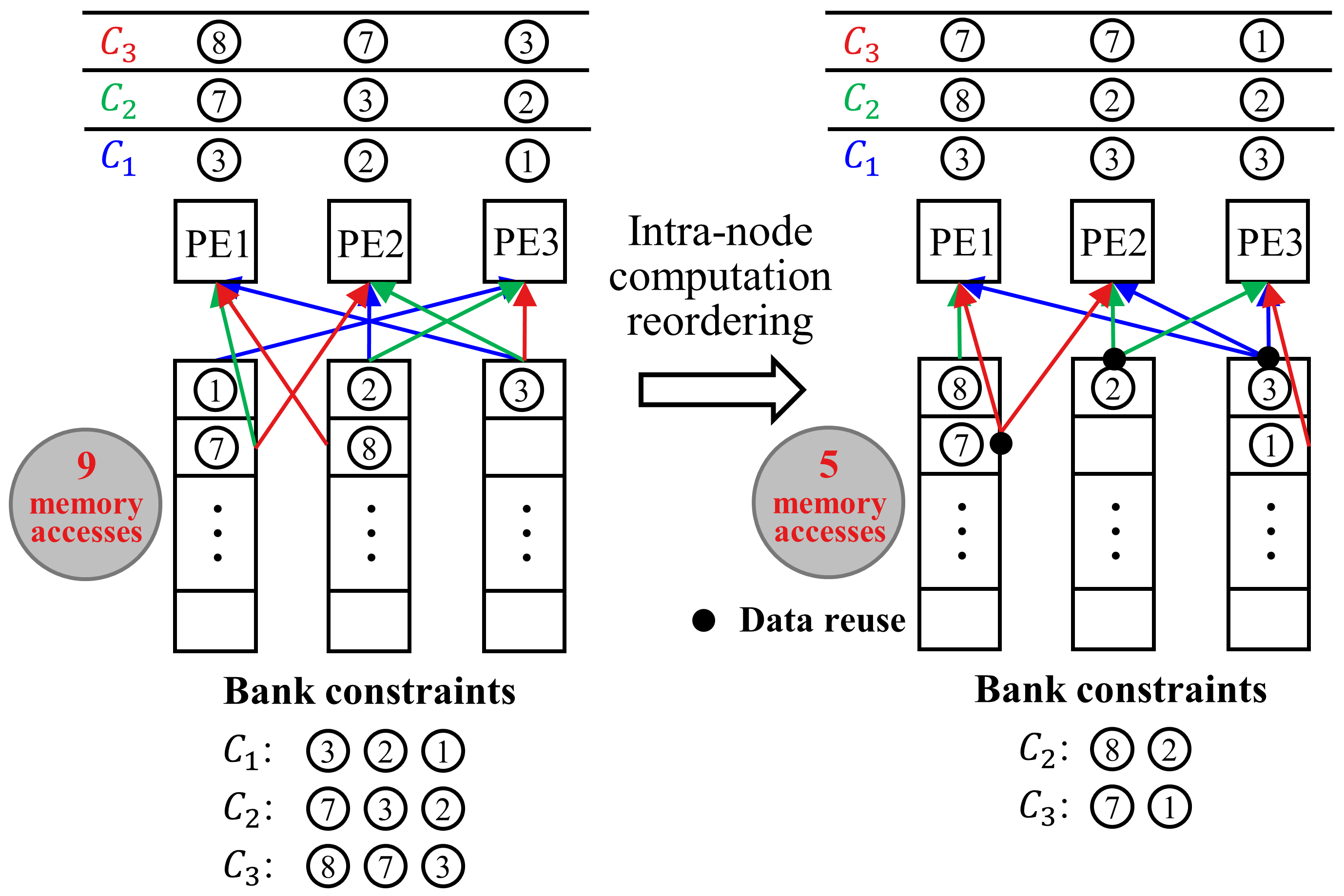}}
    \caption{Comparison of data reuse and memory access frequency with and without intra-node computation reordering.}
    \label{ICR_pic}
    \end{figure}
}

{
    \begin{algorithm}[t]
    \footnotesize
    \caption{Determine edges to be computed for PEs in a certain clock cycle}
    \label{ICR_algo}
    \SetKwInOut{Input}{Input}
    \SetKwInOut{Output}{Output}
    \Input{
    $C$: A container with $m$ sub-containers, each holding edges for computation in this cycle, where $m \leq$ number of PEs.
    }
    \Output{
    $M$: A key-value table mapping PE indices to the edges each PE computes in this cycle.
    }
    
    \tcp{Classify edges in $C$ based on their source nodes. The R-value of a category indicates the number of edges in that category.}
    $R \gets compute\_category(C)$\\
    $M \gets \phi$, $D \gets C$\\
    \While{not $D.empty()$}{
        \tcp{Find the category with the most edges in $D$.}
        $EV \gets get\_max\_category(D)$\\
        \eIf{$EV.size() \geq 2$}{
            \tcp{Find the category with the minimum R-value in $EV$ .}
            $E \gets get\_min\_R\_category(R, EV)$
        }{
            $E \gets EV[0]$
        }
        \tcp{Compute the PE index in $E$ and establish a key-value table.}
        $M\_i \gets get\_mapping(E)$\\
        $M.add(M\_i)$\\
        \tcp{Mark the sub-containers associated with each key in $M\_i$ and remove them in $D$.}
        $D\_i \gets mark\_sub\_contains(D, M\_i)$\\
        $D.remove(D\_i)$
    }
    \Return{$M$}
    \end{algorithm}
}

In each cycle, PEs may have multiple computable edges while processing the current node. Choosing which edge to compute does not affect the final result, but impacts the frequency of accessing the register banks. The traditional method selects the edge based on the ascending order of the source node ID. However, this can lead to repeated readouts of the same source node (the left subfigure of figure \ref{ICR_pic}), as edges connected to it are scattered across different positions for other nodes. We consider edges with the same source node as similar edges and propose a reordering algorithm to group similar edges within the same cycle.

The intra-node edges computation reordering algorithm (ICR) for each cycle involves the following four steps: firstly, classify similar edges and count the number in each category, denoted as the R-value (line 1 in algo. \ref{ICR_algo}). Secondly, select the category with the highest count. In case of a tie, choose the category with the smallest R-value (lines 4-9). This maximizes reuse in the current cycle and ensures nodes can be reused in subsequent cycles, as illustrated in the right subfigure of figure \ref{ICR_pic}, where PE 1 selects node 8 instead of node 7 in the second cycle, allowing node 7 to be reused in the next cycle. Thirdly, assign edges in the selected category to the corresponding PEs and remove all related edges (lines 10-13). Finally, repeat the second and third steps until all non-blocked PEs are assigned edges.

\subsection{Compiler Performance Analysis}

The time consumed by the compiler mainly comes from determining the behavior of PEs or CUs in each cycle, i.e., whether to block or compute, and whether to access $psum$ register files. The process involves identifying all edges that CUs can compute and performing the ICR algorithm. The former requires iterating over the input edges of the first unblocked node in each CU’s task list, so the computational complexity is $O(m \times p \times d)$, where $m$ is the number of cycles, $p$ is the number of CUs, and $d$ is the maximum indegree of the DAG. The latter requires selecting the edge for each CU with the smallest R-value from the remaining unallocated CUs, so the computational complexity is $O(m \times \frac{p^2}{2} \times d)$. Since $m$ scales linearly with the number of non-zeros ($nnz$) in the worst case, and $p$ is fixed for a given architecture, the computational complexity of the compiler is $O(nnz \times d)$. The memory consumption of the compiler mainly arises from the generated instructions, which scale linearly with the number of non-zeros in the worst case, so the memory complexity is $O(nnz)$.

\section{Experiments}

\subsection{Experimental Setup}

{
    \begin{table}[t]
    \caption{Synthesis Result of the Accelerator With 64 CUs}
    \begin{center}
    \begin{tabular}{l|rr|rr}
    
    \hline
     & \multicolumn{2}{c|}{Area} & \multicolumn{2}{c}{Power} \\
     & $\text{mm}^{\text{2}}$ & \% & mW & \% \\
    \hline
    Datapath:&&&& \\
    \hspace{2pt} PEs & 0.07 & 3.3 & 16.00 & 10.2 \\
    \hspace{2pt} Fifos & 0.16 & 7.7 & 28.22 & 18.1 \\
    \hspace{2pt} Pipelining registers & 0.02 & 0.8 & 6.85 & 4.4 \\
    \hspace{2pt} Input interconnect & 0.04 & 2.1 & 9.65 & 6.2 \\
    \hspace{2pt} Output interconnect & 0.04 & 2.1 & 8.36 & 5.3 \\
    Register file & 0.28 & 13.1 & 29.86 & 19.1 \\
    Control:&&&& \\
    \hspace{2pt} Control units & 0.02 & 0.9 & 5.41 & 3.5 \\
    \hspace{2pt} Multiplexers & 0.00 & 0.2 & 1.85 & 1.2 \\
    Data memory & 0.11 & 5.4 & 7.07 & 4.5 \\
    Instruction memory & 0.64 & 30.1 & 17.09 & 10.9 \\
    Stream memory & 0.72 & 34.3 & 25.86 & 16.6 \\
    \hline
     & 2.11 & & 156.21 & \\
    \hline

    \end{tabular}
    \label{synthesis}
    \end{center}
    \end{table}
}

{
    \begin{table*}[t]
    \caption{Basic Information of Partial Benchmarks}
    \begin{center}
    \begin{tabular}{|c|c|c|c|c|c|c|c|c|c|c|c|c|}
    
    \hline
       &      &   &     &        & \multicolumn{4}{c|}{CDU nodes} & Load & Peak & \multicolumn{2}{c|}{Compile time (s)}  \\
                                   \cline{6-9}                                                   \cline{12-13}
    ID & Name & N & NNZ & Binary & \multicolumn{3}{c|}{Ratio (\%)} & Edges$^{\mathrm{4}}$ & balance & throughput & DPU-v2 & This work \\
                                    \cline{6-8}
       &      &   &     & nodes  & Nodes$^{\mathrm{1}}$ & Edges$^{\mathrm{2}}$ & Levels$^{\mathrm{3}}$ & per node  & degree & (GOPS) & ($\times 10^0$) & ($\times 10^{-3}$) \\

\hline
   1 &            bp\_200 &   822 &    2874 &    4926 &   6.1 &  55.0 &  62.5 &   22 &   91.8 &  16.5 &    19.1 &    14.9 \\
\hline
   2 &           west2021 &  2021 &    6160 &   10299 &   6.8 &  38.4 &  95.8 &   11 &   17.8 &  16.1 &    65.8 &    41.1 \\
\hline
   3 &       HB\_jagmesh4 &  1440 &   22600 &   43760 &  39.2 &  88.4 &  92.6 &   33 &   22.4 &  18.6 &   163.6 &    50.2 \\
\hline
   4 &             rdb968 &   968 &   16101 &   31234 &  35.9 &  94.0 &  96.1 &   40 &   20.5 &  18.6 &   118.0 &    30.5 \\
\hline
   5 &             dw2048 &  2048 &   31909 &   61770 &  23.3 &  86.6 &  92.8 &   54 &   14.2 &  18.6 &   342.4 &    99.8 \\
\hline
   6 &        ACTIVSg2000 &  4000 &   42840 &   81680 &  10.0 &  67.6 &  91.5 &   65 &    8.8 &  18.3 &   592.3 &   248.5 \\
\hline
   7 &              cz628 &   628 &    9123 &   17618 &  57.6 &  97.4 &  98.1 &   22 &   47.3 &  18.5 &    58.3 &    15.6 \\
\hline
   8 &        bips98\_606 &  7135 &   28759 &   50383 &   3.0 &  30.8 &  83.9 &   31 &    9.8 &  16.8 &   574.6 &   304.9 \\
\hline
   9 &            nnc1374 &  1374 &   17897 &   34420 &  28.1 &  92.7 &  97.2 &   39 &   30.6 &  18.5 &   138.1 &    41.7 \\
\hline
  10 &              add20 &  2395 &    9867 &   17339 &   4.8 &  60.1 &  88.1 &   39 &   19.1 &  16.9 &    93.2 &    61.5 \\
\hline
  11 &    fpga\_trans\_01 &  1220 &    5371 &    9522 &   8.1 &  55.5 &  80.6 &   23 &   20.3 &  17.0 &    42.0 &    29.9 \\
\hline
  12 &               c-36 &  7479 &   12186 &   16893 &   0.0 &   0.0 &   0.0 &    0 &    8.2 &  13.3 &   348.4 &   117.9 \\
\hline
  13 &         circuit204 &  1020 &    8008 &   14996 &  10.8 &  58.9 &  81.5 &   37 &   24.8 &  18.0 &    60.5 &    23.7 \\
\hline
  14 &            gemat12 &  4929 &   28415 &   51901 &   5.5 &  35.9 &  81.3 &   31 &   13.4 &  17.5 &   698.1 &   255.6 \\
\hline
  15 &            bayer07 &  3268 &   26316 &   49364 &   6.9 &  29.2 &  77.8 &   29 &   21.3 &  18.0 &   919.8 &    77.1 \\
\hline
  16 &            rajat04 &  1041 &    7625 &   14209 &  13.3 &  78.0 &  88.8 &   36 &   97.6 &  17.9 &    61.4 &    37.1 \\
\hline
  17 &              add32 &  4960 &   14451 &   23942 &   1.4 &  13.6 &  75.9 &   18 &    8.2 &  15.9 &   379.7 &   173.4 \\
\hline
  18 &     fpga\_dcop\_01 &  1220 &    4303 &    7386 &   7.9 &  48.6 &  86.0 &   15 &   40.7 &  16.5 &    28.1 &    23.8 \\
\hline
  19 &           bcsstm10 &  1086 &   14546 &   28006 &  57.0 &  92.1 &  94.9 &   20 &   15.2 &  18.5 &    90.7 &    23.1 \\
\hline
  20 &            rajat19 &  1157 &    3956 &    6755 &   3.6 &  55.1 &  76.0 &   36 &  109.5 &  16.4 &    27.7 &    27.3 \\
\hline

    \multicolumn{13}{l}{$^{\mathrm{1}}$The proportion of CDU nodes among all coarse nodes.} \\
    \multicolumn{13}{l}{$^{\mathrm{2}}$The proportion of input edges for CDU nodes among all coarse edges.} \\
    \multicolumn{13}{l}{$^{\mathrm{3}}$The proportion of levels with CDU nodes among all levels.} \\
    \multicolumn{13}{l}{$^{\mathrm{4}}$The average number of input edges per CDU node.}
    
    \end{tabular}
    \label{table:benchmarks}
    \end{center}
    \end{table*}
}

The experiments compare this work with CPU, GPU, and DPU-v2 platforms. \textbf{CPU}: We use the Intel Math Kernel Library (MKL v2018.1) \cite{CPU} on a Xeon E5-2698 v4 CPU at 2.2 GHz, utilizing the $mkl\_sparse\_s\_trsv$ function. Programs are compiled with GCC v7.5.0, using the -O3 optimization flag and OpenMP v4.0.7. \textbf{GPU}: We use the cuSPARSE Library \cite{cusparse} on an RTX 2080Ti GPU at 1.35 GHz, utilizing the $cusparseScsrsv\_analysis$ function and $cusparseScsrsv\_solve$ function \cite{cusparse-2}. The $cusparseScsrsv\_analysis$ function analyzes the matrix structure and generates a dataflow that matches the GPU architecture, serving as a preprocessing or compilation step for SpTRSV. Programs are compiled with CUDA v10.2.89 compiler. Data transfer time between the host and GPU is excluded. \textbf{DPU-v2}: The accelerator runs in its default configuration with 56 PEs and 64 register files \cite{DPU-v2}.

Our accelerator is synthesized using TSMC 28 nm technology with 64 CUs. The synthesis results are shown in table \ref{synthesis}. Each CU has $x_i$ and $psum$ register files configured with 64 and 8 words. The data memory is configured with 8192 words, while both the instruction memory and stream memory are configured with 65536 words. The acceleartor is configured to a similar scale as DPU-v2, which eliminates the performance gain from the increased hardware resources. Since our PE connects the adder and multiplier in series while DPU-v2 connects in parallel, our PE can perform twice as many valid floating-point operations per clock cycle as DPU-v2. For a fair comparison, our accelerator operates at 150 MHz, half the DPU-v2 clock frequency, so that DPU-v2 and our work achieve a similar number of floating-point operations within the same time, which eliminates the performance gain from the increased clock frequency. To evaluate the performance of the proposed accelerator, we model our architecture using synopsys verilog compiled simulator (VCS) and implement the architecture with the SystemVerilog hardware description language. The performance is tested on 245 public benchmarks \cite{SuiteSparse} from various domains, such as circuit simulation and power networks. Table \ref{table:benchmarks} provides the basic information of the partial benchmarks. Our compiler is implemented in C++ 11. The running environment of the compiler is the same with GPU and DPU-v2.

{
    \begin{figure*}[htbp]
    \centerline{\includegraphics[width=1.0\linewidth]{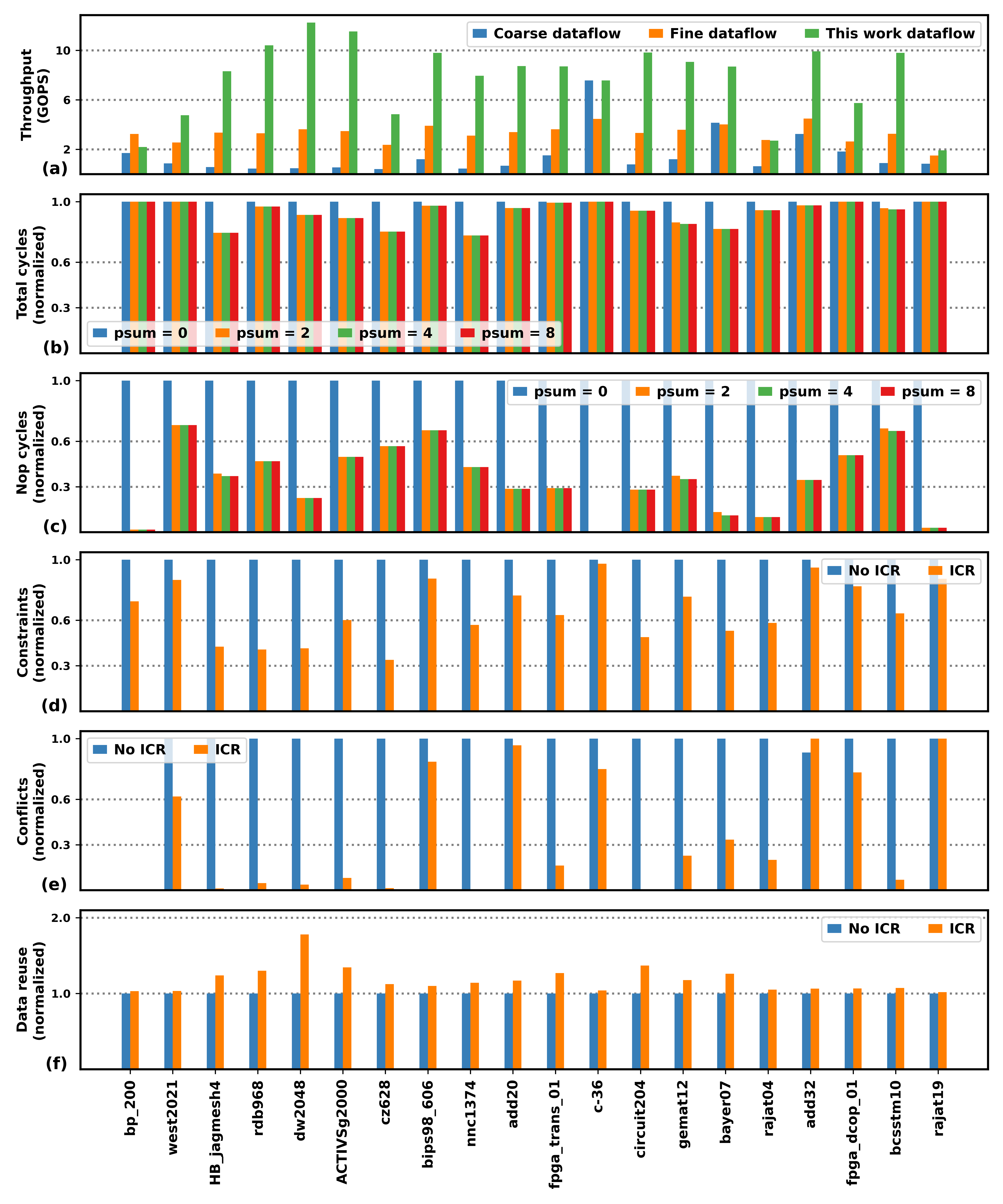}}
    \caption{
        (a) Throughput comparison of coarse, fine, and this work dataflows.
        (b) Comparison of total cycles with different $psum$ register file capacities.
        (c) Comparison of blocking cycles with different $psum$ register file capacities.
        (d) Comparison of constraints with and without intra-node edges computation reordering (ICR).
        (e) Comparison of bank conflicts with and without intra-node edges computation reordering (ICR).
        (f) Comparison of data reuse with and without intra-node edges computation reordering (ICR).
    }
    \label{experiments_dataflow_psum_icr}
    \end{figure*}
}

{
    \begin{figure}[htbp]
    \centerline{\includegraphics[width=1.0\linewidth]{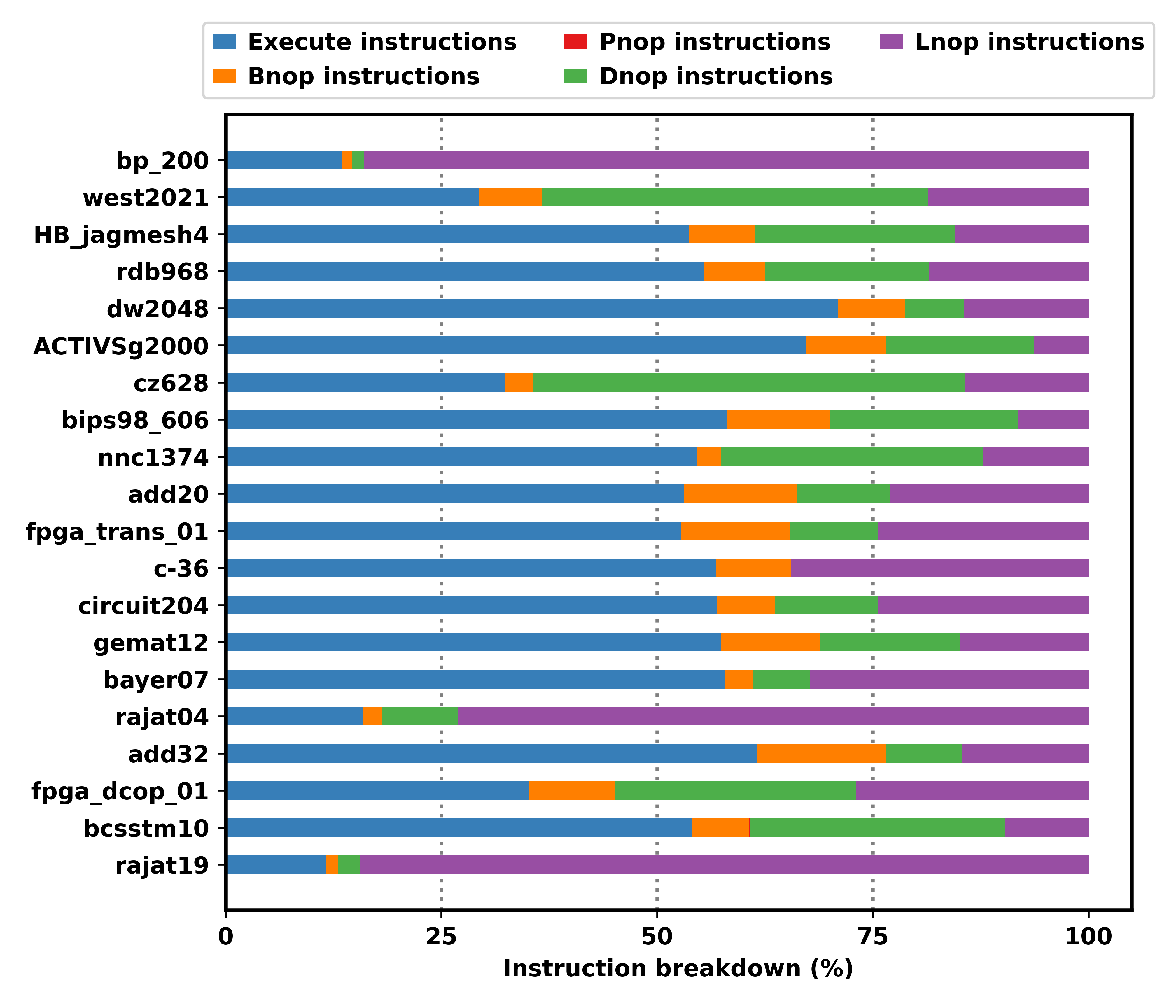}}
    \caption{
        Instruction breakdown of the partial benchmarks.
        Bnop: Blocking caused by bank conflicts.
        Pnop: Blocking caused by limited capacity of $psum$ register files.
        Dnop: Blocking caused by DAG structure or load imbalance.
        Lnop: Blocking caused by load imbalance.
    }
    \label{experiments_instruction_breakdown}
    \end{figure}
}

{
    \begin{figure*}[htbp]
    \centerline{\includegraphics[width=1.0\linewidth]{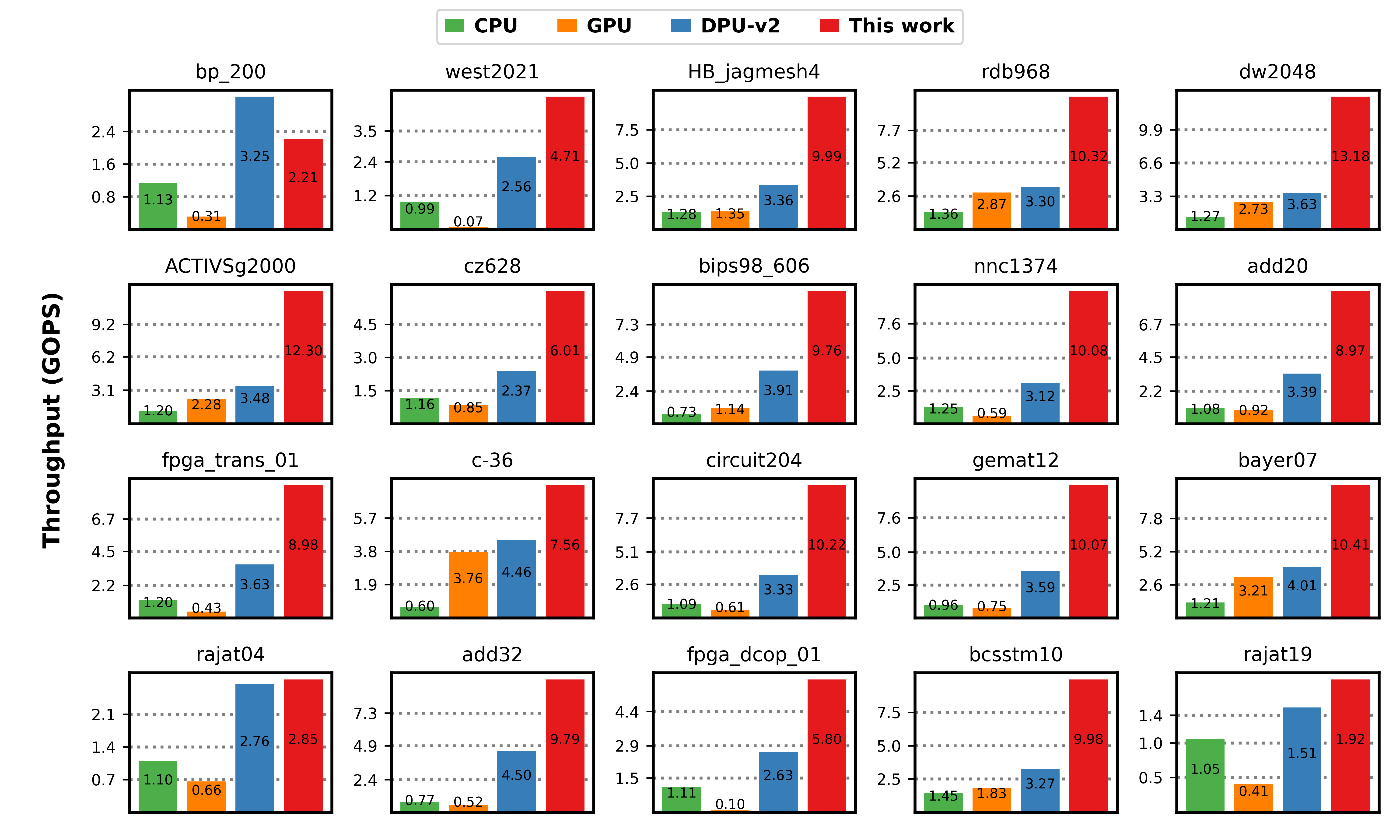}}
    \caption{
        Throughput comparison of CPU, GPU, DPU-v2, and this work on partial benchmarks.
    }
    \label{experiments_partial_throughput}
    \end{figure*}
}

\subsection{Throughput Comparison over Coarse and Fine Dataflows}

Figure \ref{experiments_dataflow_psum_icr} (a) compares the throughput of the coarse, fine, and proposed dataflows. The coarse dataflow represents the dataflow used by the synchronization-free method. For a fair comparison, the coarse dataflow is implemented on our architecture, excluding the effect of cache misses and thread synchronizations on GPUs. The fine dataflow represents the dataflow used by DPU-v2. This work dataflow does not utilize the partial sum caching mechanism. Based on the previous definition of CDU nodes, quantitative characteristics for SpTRSV-like DAGs are provided in columns 6-9 of table \ref{table:benchmarks}. The threshold is set at 20\% of our architecture's maximum parallelism (i.e., the number of CUs). Additionally, to concretely evaluate the performance of the proposed dataflow, two quantitative metrics are introduced for each benchmark. 1) Load balance degree, which is obtained by calculating the coefficient of variation for the number of input edges assigned to each CU. 2) Peak throughput, which varies across benchmarks due to differing computational demands. The peak throughput is obtained by assuming each CU performs valid computations in each cycle, as shown in equation \ref{equation:peak_throughput}, where $P$ is the number of CUs, $C$ is the clock period, $N$ is the matrix order, and $NNZ$ is the number of non-zeros of the matrix. $\frac{2\times P}{C}$ represents the peak throughput of our architecture.

{
    \begin{small}
    \begin{equation}
    peak\_throu = \frac{2 \times NNZ - N}{\frac{NNZ}{P}\times C} = \frac{2 \times P}{C}\left(1-\frac{N}{2 \times NNZ}\right)
    \label{equation:peak_throughput}
    \end{equation}
    \end{small}%
}

The coarse dataflow only performs well on few DAGs with a relatively sufficient number of independent nodes at each level, such as \textit{c-36}. However, most SpTRSV-like DAGs, such as \textit{add20}, \textit{dw2048}, and \textit{ACTIVSg2000}, contain numerous edges concentrated on CDU nodes and these nodes form long dependent chains, limiting parallelism. For the fine dataflow, coarse nodes in SpTRSV-like DAGs typically have many input edges, generating excessive intermediate fine nodes, which degrades performance. The proposed medium dataflow adopts a strategy of coarse node allocation and fine edge computation, enhancing parallelism and avoiding handling intermediate nodes, thus achieving superior performance. For certain benchmarks, such as \textit{bp\_200}, \textit{rajat04}, and \textit{rajat19}, the throughput of our dataflow is comparable to or lower than the fine dataflow. This is because the coarse node allocation strategy leads to load imbalance. In contrast, the fine dataflow decomposes coarse nodes with varying loads into fine nodes, resulting in improved load balancing. Optimizing the node allocation algorithm can mitigate load imbalance in the medium dataflow.

\subsection{Impact of Partial Sum Caching Mechanism}

Figure \ref{experiments_dataflow_psum_icr} (b) and (c) illustrate the impact of the partial sum caching mechanism on reducing blocking cycles and overall cycles under different $psum$ register files capacities. Given the wide variance in benchmark computations, the data was normalized. The results indicate that the partial sum caching mechanism reduces blocking frequency and improves overall performance. It achieves optimization with small capacities because it always prioritizes cached nodes, allowing efficiently use of the $psum$ register files. When the blocking cycles no longer change with capacity, residual blocking is attributed to the DAG structure, unresolved bank conflicts, and load imbalance. For certain benchmarks, such as matrices \textit{rajat19} and \textit{c-36}, we observe that while blocking cycles significantly decrease, the critical path is shortened less or even remains unchanged. This is because the proportion of blocking cycles is inherently small, or the reduced blocking cycles are not on the critical path, which is related to the sparse structure and coarse node allocation.

\subsection{ICR Optimization for Bank Conflicts and Data Reuse}

Figure \ref{experiments_dataflow_psum_icr} (d), (e), and (f) compare the changes in constraints, bank conflicts, and data reuse before and after applying the ICR algorithm. The ICR algorithm effectively reduces constraints and improves data reuse. Generally, reducing constraints can lower the bank conflicts because in the greedy graph coloring algorithm, fewer constraints mean more choices for nodes regarding banks allocation, thereby increasing the optimization space of these nodes and their directly or indirectly connected nodes. In few cases, such as the matrix \textit{add32}, the ICR algorithm might make bank conflicts increase. This is because the ICR algorithm focuses on the optimal computation order at the current time without considering the global. Although this can reduce the overall number of constraints, it can alter the constraint graph structure, potentially increasing constraints for some nodes. Nonetheless, such instances are rare, and in most cases, the ICR algorithm results in positive optimization.

\subsection{Instruction Breakdown}

The CUs have two states: performing valid computations or blocking, corresponding to execute and nop instructions, respectively. Figure \ref{experiments_instruction_breakdown} shows the breakdown of instructions. The nop instructions arise from four aspects: 1) bank conflicts (Bnop); 2) limited capacity of $psum$ register files (Pnop); 3) DAG structure (Dnop); 4) load imbalance caused by coarse node allocation (Dnop, Lnop). The Dnop instructions indicate that CUs have not completed their tasks, but all nodes in their task lists are blocked due to dependencies. This blocking arises from the inherent DAG structure or load imbalance. The Lnop instructions represent the blocking where CUs have completed their tasks but need to wait for other CUs to finish. This blocking is caused by load imbalance.

Thanks to the partial sum caching mechanism and the ICR algorithm, the Bnop and Pnop blocking are effectively mitigated. Blocking caused by the inherent DAG structure cannot be resolved. The remaining Blocking are due to load imbalance. Although optimizing node allocation algorithms can alleviate load imbalance, they cannot address some SpTRSV-like DAGs where a small number of coarse nodes have significantly more edges than other coarse nodes. In such cases, transforming coarse nodes into fine or medium nodes may help mitigate load imbalance. A medium node is a node that performs the same basic operations as a coarse node but has fewer input edges. Converting a coarse node into multiple fine or medium nodes introduces the challenge of handling the increased intermediate nodes but improves load balance. In fact, there is a trade-off between them, and further research is required to achieve an optimal balance.

{
    \begin{figure*}[t]
    \centerline{\includegraphics[width=0.8\linewidth]{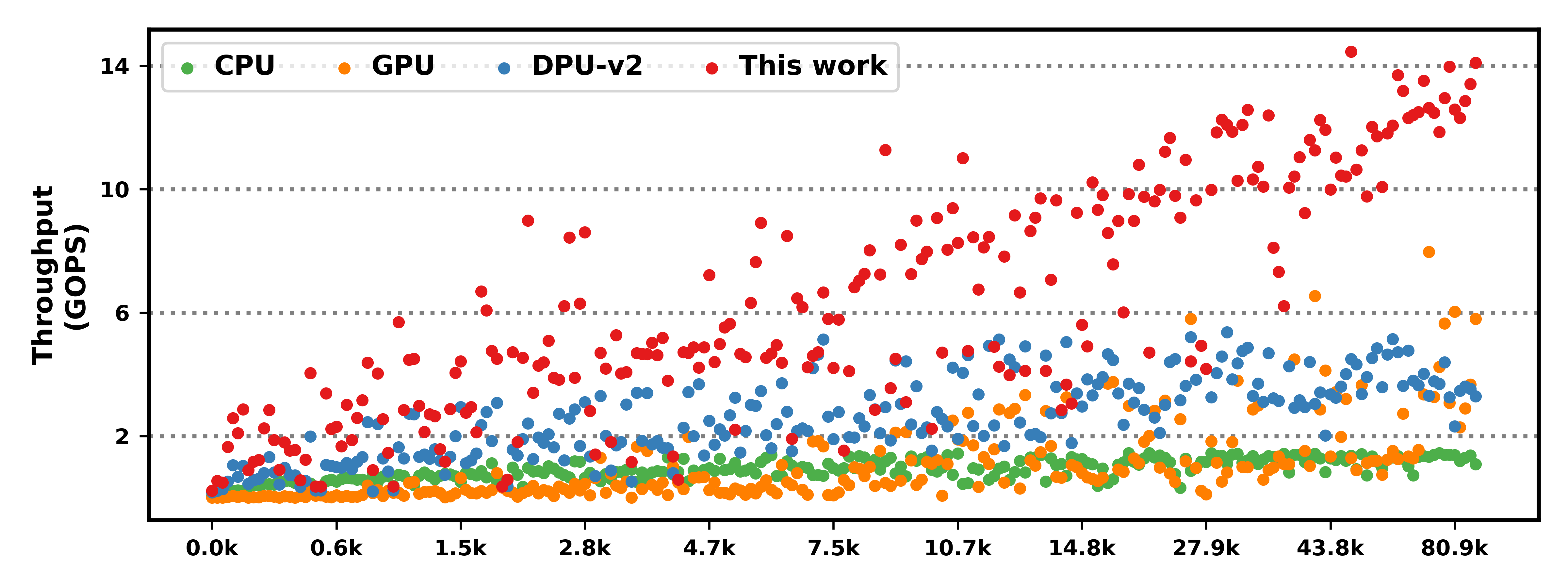}}
    \caption{
        Throughput comparison of CPU, GPU, DPU-v2, and this work on 245 benchmarks. Horizontal axis numbers represent the number of binary nodes of each benchmark, which equals the number of floating-point operations required to solve the benchmark.
    }
    \label{experiments}
    \end{figure*}
}

{
    \begin{table*}[t]
    \caption{Performance Comparison With Other Platforms on 245 SuiteSparse Benchmarks \cite{SuiteSparse}}
    \begin{center}
    \begin{tabular}{c|cccc}
    
    \hline
     & CPU & GPU & DPU-v2 & This work \\ 
    \hline
    Technology (nm)                 & 14 & 12 & 28 & 28 \\
    Frequency (MHz)                 & 2200 & 1350 & 300 & 150 \\ 
    Peak throughput (GOPS)          & 1408.0 & 13447.7 & 16.8 & 19.2 \\
    Avg. throughput (GOPS)          & 0.9 & 1.1 & 2.6 & 6.5 \\
    Speedup                         & 1$\times$ & 1.2$\times$ & 2.8$\times$ & 7.0$\times$ \\
    Power (W)                       & \textgreater50 & \textgreater50 & 0.109 & 0.156 \\
    Avg. energy efficiency (GOPS/W) & \textless0.01 & \textless0.01 & 23.9 & 41.4 \\
    Avg. compile time (s)           & - & 0.02 & 103.40 & 0.03 \\
    Avg. ratio$^{\mathrm{1}}$       & - & $1.86 \times 10^3$ & $1.62 \times 10^7$ & $1.83 \times 10^4$ \\
    \hline

    \multicolumn{5}{l}{$^{\mathrm{1}}$Average ratio of compile time to execution time.}

    \end{tabular}
    \label{performance}
    \end{center}
    \end{table*}
}

\subsection{Performance Comparison with Other Platforms}

Figure \ref{experiments_partial_throughput} shows the throughput comparison of this work with other platforms on partial benchmarks. The performance of the proposed accelerator exceeds other platforms. From figure \ref{experiments_dataflow_psum_icr} (a), it is evident that for the matrix \textit{rajat04}, the throughput of our dataflow is initially lower than DPU-v2. However, with the partial sum caching mechanism, the performance is improved and surpasses DPU-v2. Figure \ref{experiments} compares the performance of our work against other platforms across 245 benchmarks with nodes ranging from 19 to 85,392. DPU-v2 fails to produce results for 7 benchmarks due to excessive compilation time (exceeding 300 minutes). These benchmarks are \textit{odepa400}, \textit{olm1000}, \textit{lung1}, \textit{mhda416}, \textit{extr1b}, \textit{extr1}, and \textit{ex22}. The CPU and GPU exhibit low average throughput as SpTRSV disrupts both temporal and spatial locality, rendering SIMD (i.e., single instruction multiple data) or SIMT (i.e., single instruction multiple threads) functions and hardware-managed cache mechanisms inefficient \cite{challenge-cpu-gpu, challenge-cpu-gpu1, challenge-cpu-gpu2, challenge-cpu-gpu3}. Table \ref{performance} summarizes key metrics of our work and other platforms. Overall, our approach achieves average performance improvements of 7.0$\times$ (up to 27.8$\times$) and 5.8$\times$ (up to 98.8$\times$) compared to the CPU and GPU, respectively. Compared to DPU-v2, our work demonstrates an average performance improvement of 2.5$\times$ (up to 5.9$\times$) and an energy efficiency improvement of 1.7$\times$ (up to 4.1$\times$). Furthermore, the throughput of our work shows good scalability.

\subsection{Performance Comparison of Compilers}

The final two columns of table \ref{table:benchmarks} present the compilation time for DPU-v2 and this work. The significantly faster performance of our compiler can be attributed to two primary factors. First, in terms of computational complexity, DPU-v2 has a complexity of $O(T^2)$, where $T$ is the number of fine nodes, equivalent to $(2 \times nnz - n)$. Thus, the complexity of DPU-v2 is $O(nnz^2)$, whereas our compiler is $O(nnz \times d)$. Since $d$ is much smaller than $nnz$, the compilation speed of DPU-v2 is slower. The second factor is the difference in programming languages. Although running in the same environment, the DPU-v2 compiler is implemented using Python 3.7, whereas this work uses C++ 11. Table \ref{performance} shows the average compilation time for different platforms and ratios of compilation time to their respective execution time. It is worth mentioning that, while our ratio is higher than GPU, this is due to the shorter execution time of this work compared to the GPU. Therefore, in practice, the compilation time of this work is comparable to the GPU.

\section{Conclusion}


This article proposed a novel hardware accelerator based on VLIW architectures for SpTRSV. The accelerator is equipped with multiple parallel CUs, each processing a coarse node independently. The CUs communicate via input and output interconnects. The accelerator performs the medium granularity dataflow, which uses the method of coarse node allocation and fine edge computation, to ensure both high parallelism and good spatial locality. Each CU contains a $psum$ register file, which reduces blocking frequency through a partial sum caching mechanism. The accelerator uses an ICR algorithm to enhance data reuse. The accelerator is synthesized in 28-nm technology with 64 CUs and evaluated on benchmarks from the university of Florida sparse matrix collection. Experimental results show an average speedup of 7.0$\times$, 5.8$\times$, and 2.5$\times$ over state-of-the-art CPU, GPU, and specialized hardware accelerator implementations. Thus, this work demonstrates excellent performance.


\bibliographystyle{IEEEtran}

\bibliography{ref.bib}

\begin{thebibliography}{10}
\providecommand{\url}[1]{#1}
\csname url@samestyle\endcsname
\providecommand{\newblock}{\relax}
\providecommand{\bibinfo}[2]{#2}
\providecommand{\BIBentrySTDinterwordspacing}{\spaceskip=0pt\relax}
\providecommand{\BIBentryALTinterwordstretchfactor}{4}
\providecommand{\BIBentryALTinterwordspacing}{\spaceskip=\fontdimen2\font plus
\BIBentryALTinterwordstretchfactor\fontdimen3\font minus \fontdimen4\font\relax}
\providecommand{\BIBforeignlanguage}[2]{{%
\expandafter\ifx\csname l@#1\endcsname\relax
\typeout{** WARNING: IEEEtran.bst: No hyphenation pattern has been}%
\typeout{** loaded for the language `#1'. Using the pattern for}%
\typeout{** the default language instead.}%
\else
\language=\csname l@#1\endcsname
\fi
#2}}
\providecommand{\BIBdecl}{\relax}
\BIBdecl

\bibitem{DirectMethods}
T.~A. Davis, \emph{{Direct methods for sparse linear systems}}.\hskip 1em plus 0.5em minus 0.4em\relax SIAM, 2006.

\bibitem{DirectMethods-2}
T.~Wang, W.~Li, H.~Pei, Y.~Sun, Z.~Jin, and W.~Liu, ``{Accelerating sparse lu factorization with density-aware adaptive matrix multiplication for circuit simulation},'' in \emph{2023 60th ACM/IEEE Design Automation Conference (DAC)}.\hskip 1em plus 0.5em minus 0.4em\relax IEEE, 2023, pp. 1--6.

\bibitem{GPU-LU}
K.~He, S.~X.-D. Tan, H.~Wang, and G.~Shi, ``{GPU-accelerated parallel sparse LU factorization method for fast circuit analysis},'' \emph{IEEE Transactions on Very Large Scale Integration (VLSI) Systems}, vol.~24, no.~3, pp. 1140--1150, 2015.

\bibitem{GPU-LU-2}
W.-K. Lee, R.~Achar, and M.~S. Nakhla, ``{Dynamic GPU parallel sparse LU factorization for fast circuit simulation},'' \emph{IEEE Transactions on Very Large Scale Integration (VLSI) Systems}, vol.~26, no.~11, pp. 2518--2529, 2018.

\bibitem{IterativeMethods}
Y.~Saad, \emph{{Iterative methods for sparse linear systems}}.\hskip 1em plus 0.5em minus 0.4em\relax SIAM, 2003.

\bibitem{IterativeMethods-2}
H.~Anzt, E.~Chow, and J.~Dongarra, ``{Iterative sparse triangular solves for preconditioning},'' in \emph{Euro-Par 2015: Parallel Processing: 21st International Conference on Parallel and Distributed Computing, Vienna, Austria, August 24-28, 2015, Proceedings 21}.\hskip 1em plus 0.5em minus 0.4em\relax Springer, 2015, pp. 650--661.

\bibitem{SpMV}
N.~Jao, A.~K. Ramanathan, J.~Sampson, and V.~Narayanan, ``{Sparse Vector-Matrix Multiplication Acceleration in Diode-Selected Crossbars},'' \emph{IEEE Transactions on Very Large Scale Integration (VLSI) Systems}, vol.~29, no.~12, pp. 2186--2196, 2021.

\bibitem{SpMM}
E.~B. Tavakoli, M.~Riera, M.~H. Quraishi, and F.~Ren, ``{FSpGEMM: A Framework for Accelerating Sparse General Matrix--Matrix Multiplication Using Gustavson’s Algorithm on FPGAs},'' \emph{IEEE Transactions on Very Large Scale Integration (VLSI) Systems}, 2024.

\bibitem{PGT-SOLVER}
T.~Yu and M.~D. Wong, ``{PGT\_SOLVER: An efficient solver for power grid transient analysis},'' in \emph{2012 IEEE/ACM International Conference on Computer-Aided Design (ICCAD)}.\hskip 1em plus 0.5em minus 0.4em\relax IEEE, 2012, pp. 647--652.

\bibitem{CircuitSimulation}
F.~N. Najm, \emph{{Circuit simulation}}.\hskip 1em plus 0.5em minus 0.4em\relax John Wiley \& Sons, 2010.

\bibitem{ParallelDirect}
X.~Chen, Y.~Wang, and H.~Yang, \emph{{Parallel sparse direct solver for integrated circuit simulation}}.\hskip 1em plus 0.5em minus 0.4em\relax Springer, 2017.

\bibitem{GPU-SpTRSV-kernel}
M.~Freire, J.~Ferrand, F.~Seveso, E.~Dufrechou, and P.~Ezzatti, ``{A GPU method for the analysis stage of the SPTRSV kernel},'' \emph{The Journal of Supercomputing}, vol.~79, no.~13, pp. 15\,051--15\,078, 2023.

\bibitem{level-scheduling}
E.~Anderson and Y.~Saad, ``{Solving sparse triangular linear systems on parallel computers},'' \emph{International Journal of High Speed Computing}, vol.~1, no.~01, pp. 73--95, 1989.

\bibitem{level-scheduling-2}
J.~H. Saltz, ``{Aggregation methods for solving sparse triangular systems on multiprocessors},'' \emph{SIAM journal on scientific and statistical computing}, vol.~11, no.~1, pp. 123--144, 1990.

\bibitem{sync-free-1}
W.~Liu, A.~Li, J.~Hogg, I.~S. Duff, and B.~Vinter, ``{A synchronization-free algorithm for parallel sparse triangular solves},'' in \emph{Euro-Par 2016: Parallel Processing: 22nd International Conference on Parallel and Distributed Computing, Grenoble, France, August 24-26, 2016, Proceedings 22}.\hskip 1em plus 0.5em minus 0.4em\relax Springer, 2016, pp. 617--630.

\bibitem{sync-free-2}
W.~Liu, A.~Li, J.~D. Hogg, I.~S. Duff, and B.~Vinter, ``{Fast synchronization-free algorithms for parallel sparse triangular solves with multiple right-hand sides},'' \emph{Concurrency and Computation: Practice and Experience}, vol.~29, no.~21, p. e4244, 2017.

\bibitem{sptrsv-spmv-1}
N.~Ahmad, B.~Yilmaz, and D.~Unat, ``{A split execution model for sptrsv},'' \emph{IEEE Transactions on Parallel and Distributed Systems}, vol.~32, no.~11, pp. 2809--2822, 2021.

\bibitem{sptrsv-spmv-2}
Z.~Lu, Y.~Niu, and W.~Liu, ``{Efficient block algorithms for parallel sparse triangular solve},'' in \emph{Proceedings of the 49th International Conference on Parallel Processing}, 2020, pp. 1--11.

\bibitem{TileSpTRSV}
Z.~Lu and W.~Liu, ``{TileSpTRSV: a tiled algorithm for parallel sparse triangular solve on GPUs},'' \emph{CCF Transactions on High Performance Computing}, vol.~5, no.~2, pp. 129--143, 2023.

\bibitem{sptrsv-rewrite}
B.~Y{\i}lmaz and A.~F. Y{\i}ld{\i}z, ``{A Graph Transformation Strategy for Optimizing SpTRSV},'' \emph{arXiv preprint arXiv:2206.05843}, 2022.

\bibitem{sptrsv-rewrite-2}
B.~Y{\i}lmaz, ``{A novel graph transformation strategy for optimizing SpTRSV on CPUs},'' \emph{Concurrency and Computation: Practice and Experience}, vol.~35, no.~24, p. e7761, 2023.

\bibitem{DPU-v2}
N.~Shah, W.~Meert, and M.~Verhelst, ``{DPU-v2: Energy-efficient execution of irregular directed acyclic graphs},'' in \emph{2022 55th IEEE/ACM International Symposium on Microarchitecture (MICRO)}.\hskip 1em plus 0.5em minus 0.4em\relax IEEE, 2022, pp. 1288--1307.

\bibitem{DPU}
N.~Shah, L.~I.~G. Olascoaga, S.~Zhao, W.~Meert, and M.~Verhelst, ``{DPU: DAG processing unit for irregular graphs with precision-scalable posit arithmetic in 28 nm},'' \emph{IEEE Journal of Solid-State Circuits}, vol.~57, no.~8, pp. 2586--2596, 2021.

\bibitem{DPU-book}
N.~Shah, W.~Meert, and M.~Verhelst, \emph{{Efficient Execution of Irregular Dataflow Graphs: Hardware/Software Co-optimization for Probabilistic AI and Sparse Linear Algebra}}.\hskip 1em plus 0.5em minus 0.4em\relax Springer Nature, 2023.

\bibitem{CSR-CSC}
D.~Langr and P.~Tvrdik, ``{Evaluation criteria for sparse matrix storage formats},'' \emph{IEEE Transactions on parallel and distributed systems}, vol.~27, no.~2, pp. 428--440, 2015.

\bibitem{FPGA-SpTRSV}
F.~Favaro, E.~Dufrechou, P.~Ezzatti, and J.~P. Oliver, ``{Exploring FPGA optimizations to compute sparse numerical linear algebra kernels},'' in \emph{Applied Reconfigurable Computing. Architectures, Tools, and Applications: 16th International Symposium, ARC 2020, Toledo, Spain, April 1--3, 2020, Proceedings 16}.\hskip 1em plus 0.5em minus 0.4em\relax Springer, 2020, pp. 258--268.

\bibitem{FPGA-SpTRSV-2}
Z.~He, L.~Song, R.~F. Lucas, and J.~Cong, ``{LevelST: Stream-based Accelerator for Sparse Triangular Solver},'' in \emph{Proceedings of the 2024 ACM/SIGDA International Symposium on Field Programmable Gate Arrays}, 2024, pp. 67--77.

\bibitem{challenge-cpu-gpu}
K.~Lu, Z.~Li, L.~Liu, J.~Wang, S.~Yin, and S.~Wei, ``{Redesk: A reconfigurable dataflow engine for sparse kernels on heterogeneous platforms},'' in \emph{2019 IEEE/ACM International Conference on Computer-Aided Design (ICCAD)}.\hskip 1em plus 0.5em minus 0.4em\relax IEEE, 2019, pp. 1--8.

\bibitem{challenge-cpu-gpu1}
S.~Li, D.~Liu, and W.~Liu, ``{Optimized data reuse via reordering for sparse matrix-vector multiplication on fpgas},'' in \emph{2021 IEEE/ACM International Conference On Computer Aided Design (ICCAD)}.\hskip 1em plus 0.5em minus 0.4em\relax IEEE, 2021, pp. 1--9.

\bibitem{challenge-cpu-gpu2}
B.~Liu and D.~Liu, ``{Towards high-bandwidth-Utilization SpMV on FPGAs via partial vector duplication},'' in \emph{Proceedings of the 28th Asia and South Pacific Design Automation Conference}, 2023, pp. 33--38.

\bibitem{challenge-cpu-gpu3}
E.~Yi, Y.~Duan, Y.~Bai, K.~Zhao, Z.~Jin, and W.~Liu, ``{Cuper: Customized Dataflow and Perceptual Decoding for Sparse Matrix-Vector Multiplication on HBM-Equipped FPGAs},'' in \emph{2024 Design, Automation \& Test in Europe Conference \& Exhibition (DATE)}.\hskip 1em plus 0.5em minus 0.4em\relax IEEE, 2024, pp. 1--6.

\bibitem{VLIW-1}
A.~Capitanio, N.~Dutt, and A.~Nicolau, ``{Partitioned register files for VLIWs: A preliminary analysis of tradeoffs},'' \emph{ACM SIGMICRO Newsletter}, vol.~23, no. 1-2, pp. 292--300, 1992.

\bibitem{VLIW-2}
J.~A. Fisher, P.~Faraboschi, and C.~Young, \emph{{Embedded computing: a VLIW approach to architecture, compilers and tools}}.\hskip 1em plus 0.5em minus 0.4em\relax Elsevier, 2005.

\bibitem{VLIW-3}
A.~K. Jones, R.~Hoare, D.~Kusic, J.~Fazekas, and J.~Foster, ``{An FPGA-based VLIW processor with custom hardware execution},'' in \emph{Proceedings of the 2005 ACM/SIGDA 13th international symposium on Field-programmable gate arrays}, 2005, pp. 107--117.

\bibitem{VLIW-4}
M.~Lam, ``{Software pipelining: An effective scheduling technique for VLIW machines},'' in \emph{Proceedings of the ACM SIGPLAN 1988 conference on Programming Language design and Implementation}, 1988, pp. 318--328.

\bibitem{VLIW-5}
D.~Sabena, M.~S. Reorda, and L.~Sterpone, ``{On the automatic generation of optimized software-based self-test programs for VLIW processors},'' \emph{IEEE Transactions on Very Large Scale Integration (VLSI) Systems}, vol.~22, no.~4, pp. 813--823, 2013.

\bibitem{sum-product}
H.~Poon and P.~Domingos, ``{Sum-product networks: A new deep architecture},'' in \emph{2011 IEEE International Conference on Computer Vision Workshops (ICCV Workshops)}.\hskip 1em plus 0.5em minus 0.4em\relax IEEE, 2011, pp. 689--690.

\bibitem{sum-product-2}
Y.~Choi, A.~Vergari, and G.~Van~den Broeck, ``{Probabilistic circuits: A unifying framework for tractable probabilistic models},'' \emph{UCLA. URL: \url{http://starai.cs.ucla.edu/papers/ProbCirc20.pdf}}, p.~6, 2020.

\bibitem{CPU}
Intel, ``{Intel Math Kernel Library},'' \url{https://www.intel.com/content/www/us/en/developer/tools/oneapi/onemkl.html}, 2018.

\bibitem{cusparse}
M.~Naumov, L.~Chien, P.~Vandermersch, and U.~Kapasi, ``{Cusparse library},'' in \emph{GPU Technology Conference}, 2010.

\bibitem{cusparse-2}
M.~Naumov, ``{Parallel solution of sparse triangular linear systems in the preconditioned iterative methods on the GPU},'' \emph{NVIDIA Corp., Westford, MA, USA, Tech. Rep. NVR-2011}, vol.~1, 2011.

\bibitem{SuiteSparse}
T.~A. Davis and Y.~Hu, ``{The University of Florida sparse matrix collection},'' \emph{ACM Transactions on Mathematical Software (TOMS)}, vol.~38, no.~1, pp. 1--25, 2011.

\end{thebibliography}

\end{document}